\documentclass{egpubl}
\usepackage{eg2026}
 
\ConferencePaper

\CGFStandardLicense

\usepackage[T1]{fontenc}
\usepackage{dfadobe}  

\usepackage{cite}
\BibtexOrBiblatex
\electronicVersion
\PrintedOrElectronic
\ifpdf \usepackage[pdftex]{graphicx} \pdfcompresslevel=9
\else \usepackage[dvips]{graphicx} \fi

\usepackage{egweblnk} 
\usepackage{lineno}

\usepackage{xcolor}
\usepackage{mathtools}
\usepackage{xspace}
\usepackage[ruled]{algorithm2e}
\usepackage{wrapfig}
\usepackage[normalem]{ulem}

\title{TABI: Tight and Balanced Interactive Atlas Packing}

\author[Floria Gu, Nicholas Vining, and Alla Sheffer]
{\parbox{\textwidth}{\centering F. Gu$^{1}$\orcid{0009-0002-0625-3247}, N. Vining$^{2}$\orcid{0009-0008-4369-7852}, A. Sheffer$^{1}$\orcid{0000-0001-9251-3716}}
        \\
{\parbox{\textwidth}{\centering $^1$University of British Columbia, Canada\\
$^2$NVIDIA, Canada\\
}
}
}

\newenvironment{parWithWrapFigure} %
{\begingroup
\setlength{\columnsep}{1em}%
\setlength{\intextsep}{0em}%
\setlength{\arraycolsep}{0pt}} %
{

\endgroup}

\newcommand{\FLIP}{\protect\reflectbox{F}LIP\xspace}

\begin{document}

\teaser{
\includegraphics[width=\linewidth]{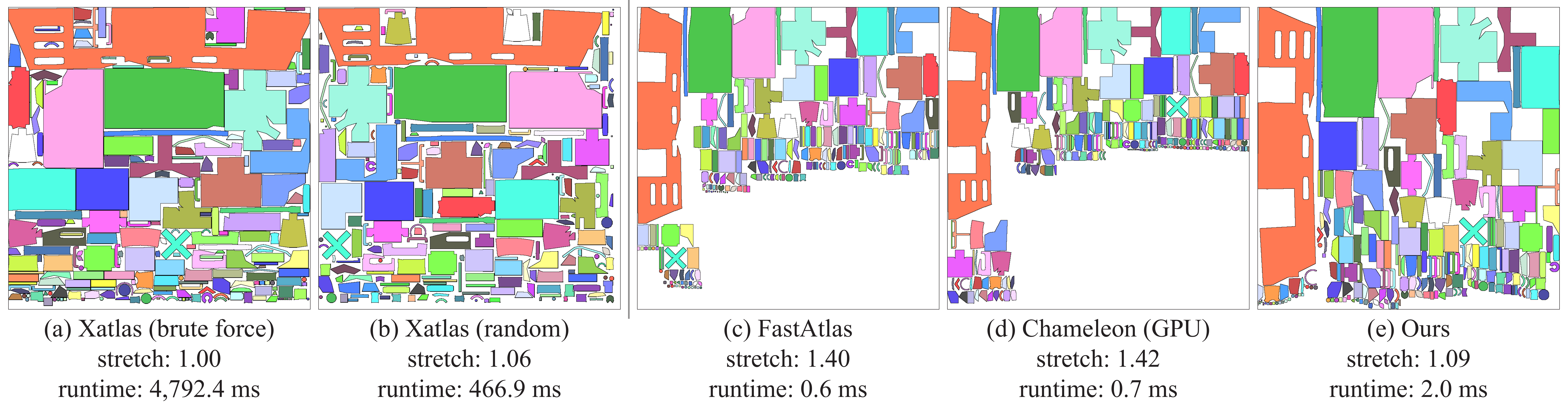}
\centering
\caption{Offline packing methods, such as Xatlas \cite{xatlas} (a,b), generate tight, high-quality packings, only minimally downscaling input charts. However, their runtimes are orders of magnitude higher than the threshold for interactive performance. Interactive packing methods such as FastAtlas \cite{vining2025fastatlas} (c) and GPU Chameleon \cite{igarashi2001adaptive} (d) generate packings which leave empty spaces between and underneath charts, requiring significantly larger chart downscaling to fit the charts into fixed size atlases. Our TABI packing algorithm (e) explicitly targets both tightness and balance, and achieves packing scale (as measured by $L^2$ stretch) nearly comparable to offline packing methods within an interactive runtime budget.}
\label{fig:teaser}
}

\maketitle
\begin{abstract}
Atlas packing is a key step in many computer graphics applications. Packing algorithms seek to arrange a set of charts within a fixed-size atlas with as little downscaling as possible.  Many packing applications such as content creation tools, dynamic atlas generation for video games, and texture space shading require on-the-fly interactive atlas packing. Unfortunately, while many methods have been developed for generating tight high-quality packings, they are designed for {\em offline} settings and have running times two or more orders of magnitude greater than what is required for interactive performance. While real-time GPU packing methods exist, they significantly downscale packed charts compared to offline methods. We introduce a GPU packing method that targets interactive speeds, provides packing quality approaching that of offline methods, and supports flexible user control over the tradeoff between performance and quality. We observe that current real-time packing methods leave large gaps between charts and often produce asymmetric, or poorly balanced, packings. These artifacts dramatically degrade packing quality. Our {\em Tight} And {\em Balanced} method eliminates these artifacts while retaining {\em Interactive} performance. TABI generates tight packings by compacting empty space between irregularly shaped charts both horizontally and vertically, using two approximations of chart shape that support efficient parallel processing. We balance packing outputs by automatically adjusting atlas row widths and orientations to accommodate varying chart heights. We show that our method significantly reduces chart downscaling compared to existing interactive methods while remaining orders of magnitude faster than offline alternatives.

\end{abstract}

\clearpage

\section{Introduction}

\begin{figure}
\includegraphics[width=\linewidth]{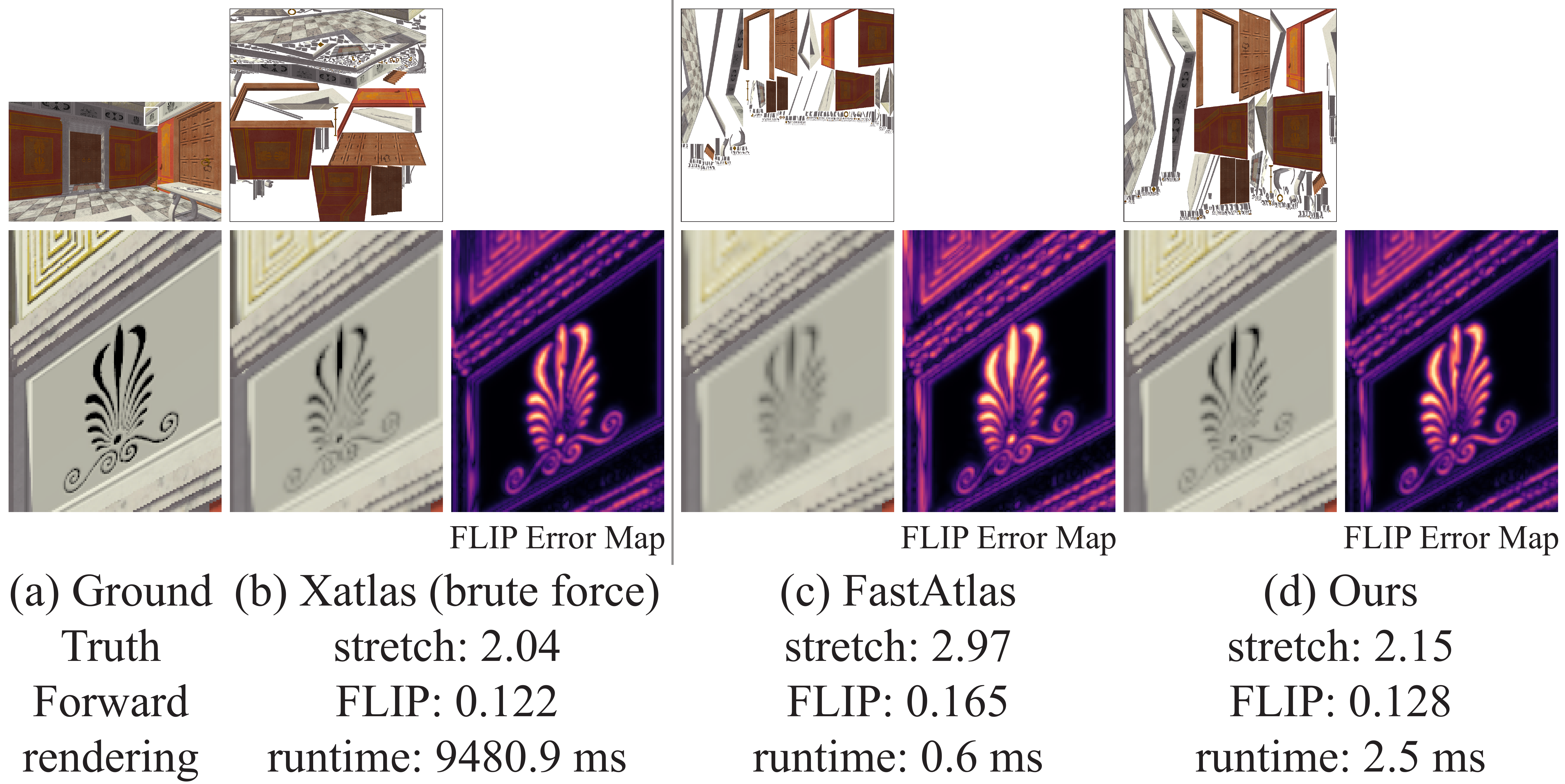}
\caption{Atlas packing scale directly impacts visual quality in downstream applications. On the reference frame shown in (a), texture-space shaded renders generated using FastAtlas (c) exhibit significant blur as reflected in the \FLIP error and error maps (black low error; pink to yellow high). This is caused by low packing scale, as measured by the $L^2$ stretch. Renders generated using our method (d) more faithfully preserve the details of the frieze and closely approach the quality of renders generated using Xatlas (b), an offline packing method. All methods use a 1K $\times$ 1K atlas.}
\label{fig:impact}
\end{figure}

Computer graphics applications routinely use 2D atlases to store surface signals, such as albedo or normal maps. This process involves cutting surfaces into charts, parameterizing these charts in 2D, and finally packing the 2D charts into an atlas.  Our work focuses on this final packing stage. 
Packing methods typically aim to pack a set of parameterized 2D charts, whose dimensions are determined by the resolution of the stored signal, into a fixed size atlas. While this process often requires scaling the charts down to fit them into the atlas, packing methods seek to minimize downscaling to preserve the input signal as accurately as possible (Fig.~\ref{fig:impact}).
For many applications in games and other settings (e.g. lightmapping \cite{arvo1986backward,SourceEngine}, surface radiance caching \cite{wright2022lumen}, texture-space shading \cite{baker2012rock}), the content being atlased is unknown until run-time. Thus, a packing must be computed at interactive rates (under 15 milliseconds when targeting 60 FPS). We propose a new packing algorithm which achieves interactive performance while drastically reducing the amount of downscaling compared to prior interactive alternatives. 

Computing optimal packings (ones that require the smallest amount of downscaling) is known to be NP-hard \cite{milenkovic1999}. Existing methods use a range of heuristics to balance runtime against downscaling (Sec.~\ref{sec:related}). Some downscaling may be unavoidable, for instance if the area of the charts exceeds the area of the atlas; but intuitively, avoidable downscaling strongly and inversely correlates to the amount of space in the packed atlas that is not occupied by charts, as less ``wasted'' area reflects less unnecessary downscaling. 

Offline packing methods (Sec.~\ref{sec:related}) produce atlases with relatively little wasted space, but at the cost of long run times; for example, the industry standard Xatlas packing library \cite{xatlas}, using its faster ``random'' setting, requires nearly 500 milliseconds to pack the input in Fig. \ref{fig:teaser}b (214 charts), and requires 2.5 {\em seconds} or more to pack the input in Fig~\ref{fig:results_tss}b (1,572 charts). Interactive performance requires speeds which are at least two orders of magnitude faster than this. Methods that achieve such performance \cite{Neff2022MSA, vining2025fastatlas} do so by taking advantage of GPU parallelism. Their performance improvement comes at the expense of a dramatic increase in downscaling (Fig~\ref{fig:teaser}c, measured using the $L^2$ stretch of the mapping from the downscaled atlased charts to their input counterparts \cite{sander2001texture,vining2025fastatlas}; a stretch of 1 is ideal, indicating zero downscaling, while increased stretch reflects more downscaling). 
 
Our GPU-based packing algorithm is 464 times faster on average than the best offline alternative, while dramatically reducing stretch compared to prior interactive alternatives (Fig~\ref{fig:teaser}e).
We achieve this reduction in stretch by identifying and addressing the two primary sources of wasted space in the best performing interactive method, FastAtlas \cite{vining2025fastatlas}: lack of {\em tightness} and {\em imbalance} (Sec~\ref{sec:explain}).
FastAtlas, and the Chameleon \cite{igarashi2001adaptive} method that inspired it (Fig~\ref{fig:teaser}d), pack the bounding rectangles of input charts, leaving the space between charts and their bounding boxes wasted. Additionally, these methods also often generate height-imbalanced packings that leave large areas at the bottom of the atlas unoccupied by any charts (see the empty spaces on the bottom right of the atlases in Fig~\ref{fig:teaser}cd). Jointly, these two factors lead to significantly larger chart downscaling compared to offline alternatives.  Our {\textbf T}ight {\textbf A}nd {\textbf B}alanced {\textbf I}nteractive (\textbf{TABI}) method reduces downscaling by decreasing the amount of space wasted around each chart, and by better balancing atlas chart placement (Sec~\ref{sec:algo}, Fig~\ref{fig:teaser}e). TABI produces tighter packings at interactive speeds by employing approximations of chart shape that can be efficiently used to compact empty space between charts both horizontally and vertically (Sec~\ref{sec:tight}). 
We improve balance by dynamically adjusting atlas row widths and directions to account for height differences between charts (Sec~\ref{sec:balance}). Our method takes advantage of GPU parallelism wherever possible, and permits users to adjust the performance/quality trade-off for their needs. It supports fixed width gutters \cite{gonzalez2009continuity}, necessary for mipmapping, seam management, and higher-order filtering; and can be used both in settings where charts can be freely rotated to minimize downscaling and in those where only multiple-of-$90^{\circ}$ rotations are allowed. 

We validate the performance of our method by applying it to 8,386 differently sourced chart sets, packing each set at two different target atlas sizes (Sec~\ref{sec:results}).
We test both chart sets originating from UV-unwrapped 3D models (455 inputs), and texture-space shading chart sets produced by \cite{vining2025fastatlas} (7,931 inputs); the latter includes inputs with tens of thousands of charts (the largest containing 44,389 charts), all robustly handled by our method. 
We compare our runtimes and stretch against interactive packing alternatives as well as the industry-standard Xatlas library \cite{xatlas}, a highly optimized implementation of the standard Tetris offline packing algorithm \cite{levy2002least}. Relative to the best prior interactive method, TABI significantly reduces chart downscaling (narrows the gap in atlas stretch, compared to offline methods, by 48\% on average) and thus enables the generation of higher quality textured atlases. While TABI is slower than methods like FastAtlas, it remains fully interactive (average runtime 5.11ms, runtime for atlases with over 10,000 charts 10.79ms). TABI runtimes are 464 times faster on average than those of the fastest offline alternative.

\section{Related Work} \label{sec:related}

\textit{Atlas Generation.}
Atlas generation for storing surface signals, such as textures or normal maps, is a well-researched problem~\cite{sheffer2007mesh}. Atlassing typically consists of three steps, the first two of which are sometimes intertwined \cite{sorkine2002bounded,li2018optcuts}: segmenting input surfaces into charts \cite{julius2005d,zhou2004isocharts}; parameterizing these charts into 2D \cite{levy2002least,sheffer2005abf++}; and finally, packing the parameterized 2D charts into an atlas \cite{levy2002least,igarashi2001adaptive}. Our work focuses on the final packing step of this process, and makes no assumptions about the source, size, shape or count of 2D charts in each input. 
 
\textit{Atlas Packing.}
Packing 2D charts into atlases as tightly as possible is a well-researched NP-hard problem; see \cite{guo2022} for a recent survey. Traditional packing algorithms seek near-optimal solutions and solve for exact polygon placement using tools such as the no-fit polygon \cite{bennell2008}, then run heuristic search algorithms to evaluate many different placement orders. These methods are computationally infeasible for computer graphics applications \cite{yang2023learningbased}, where typical inputs can consist of hundreds or thousands of arbitrarily-shaped charts. 

Algorithms for practical graphics applications typically first sort charts in decreasing order of size, then place them into the atlas one at a time. Chameleon\cite{igarashi2001adaptive} approximates charts by their bounding rectangles, folds them into rows, places rows into the atlas in alternating directions, and pushes each row up against the row above it (Fig~\ref{fig:fa_igarashi}b). Chameleon can be parallelized and is amenable to GPU implementation (\cite{vining2025fastatlas}, Sec.~\ref{sec:background}), allowing for interactive performance. 

The ``Tetris'' packing method of \cite{levy2002least} and its extensions \cite{sander2003multichart, zhou2004isocharts, noll2011efficient} represent both the charts and the current packing as rasterized images, and search for valid (non-overlapping) positions and orientations within the rasterized atlas. The best candidate position for each chart is selected according to a heuristic score function.
Xatlas \cite{xatlas} is a highly optimized implementation of Tetris packing that is widely used in industry; it supports both a slower ``brute force'' packing (Fig~\ref{fig:teaser}a) and a faster pseudo-randomized ``random'' one (Fig~\ref{fig:teaser}b). While much slower than CPU or GPU Chameleon, these methods require significantly less downscaling and are widely used for applications where runtime is a lesser concern. Tetris packing methods are inherently sequential; each inserted chart's location depends on the final locations of all prior charts, and thus they are not amenable to parallelization \cite{vining2025fastatlas}. We aim for interactive runtimes orders of magnitude faster than those of Xatlas, while approaching its output quality (Fig~\ref{fig:teaser}e, Sec.~\ref{sec:results}).

\textit{Learning-Based Packing.}
Recent works have explored learning-based approaches to packing \cite{fang2023hybrid, yang2023learningbased,xue2023learning}, and have shown packing efficiency improvements (lower stretch) over Tetris packing approaches. However, these methods are even slower than Tetris-type approaches, making them unsuitable for interactive applications. For instance, \cite{yang2023learningbased} requires several minutes to pack their largest chart set (784 charts).

\textit{Fixed Size Atlases and Gutters.}
Most atlassing applications target a fixed atlas size; however, most parameterization methods operate in a continuous space. During the packing step, the charts they produce must be scaled to fit the target atlas. A naive approach would be to compute either a continuous-space packing or a very high-resolution raster one, then scale the packed atlas to fit into the fixed size target atlas. When using a texture atlas, however, most applications use bilinear or other sampling strategies that require the charts in the atlas to have fixed width (1px or more) empty gutters around them \cite{purnomo2004seamless} (i.e. a distance between charts of 2px or more). Uniformly scaling an existing packing {\em cannot} guarantee that each chart has the required gutters around it, as scaling and rounding may produce inter-chart gaps narrower than double the gutter width. Scaling can also cause charts to overlap when two previously disjoint chart boundary points are quantized to the same location. Strictly avoiding overlaps and preserving gutters requires scaling charts by a pre-determined factor first, then packing the scaled charts in a raster space whose dimensions match the target atlas size. However, using an unnecessarily low scale factor is undesirable as it may reduce the quality of the signal stored.
Practical atlassing methods \cite{noll2011efficient,sander2003multichart,vining2025fastatlas} therefore search for the scaling factor that minimizes downscaling by packing atlases with different factors and selecting the largest one that produces a valid output. We adopt this framework for evaluating both our method and alternative algorithms. We enforce 1px gutters for both our and alternative methods in all experiments. 

\begin{figure}
\includegraphics[width=\linewidth]{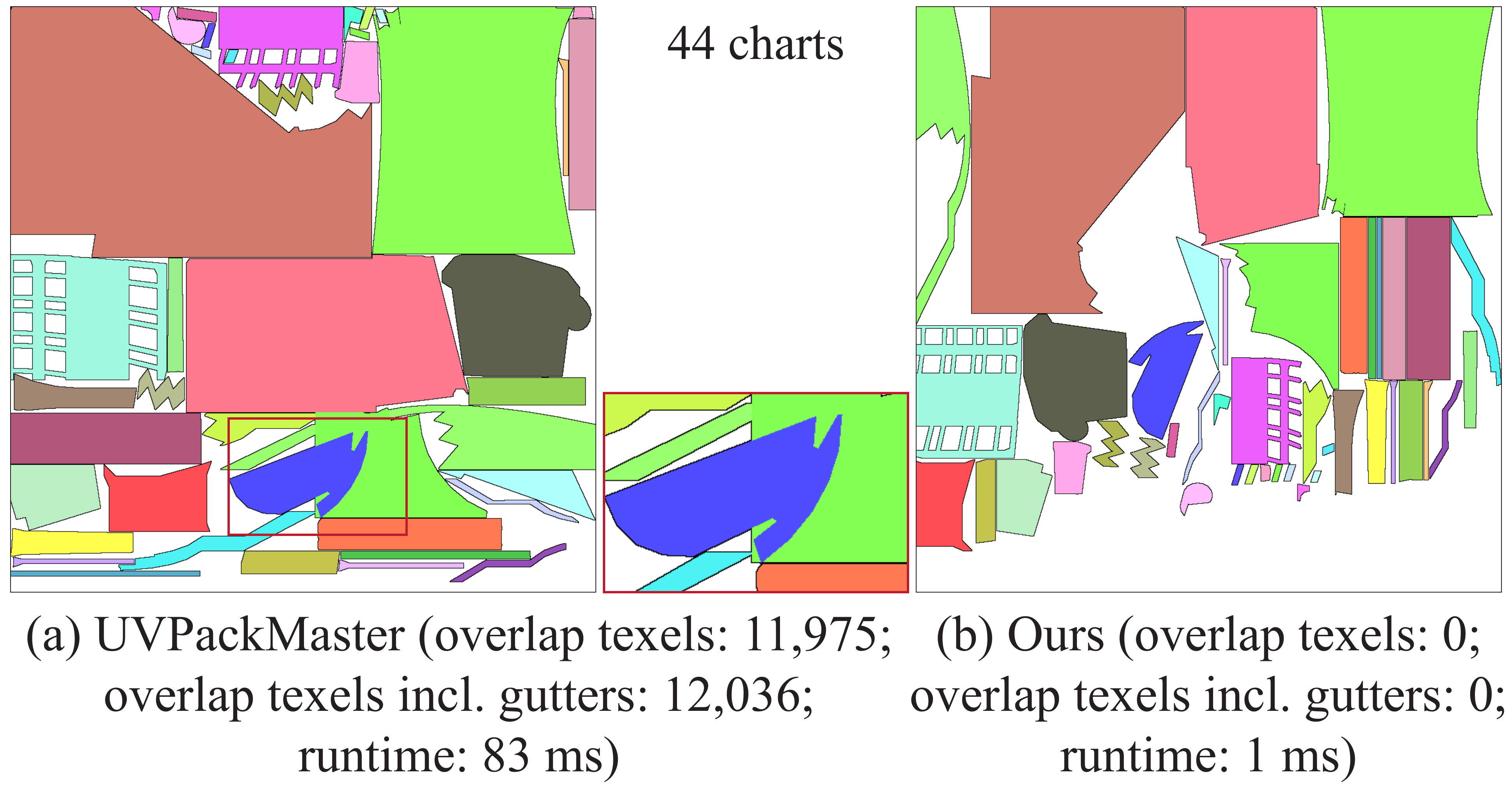}
\caption{The GPU UVPackMaster packer \cite{uvpackmaster} outputs have severe inter-chart overlaps on even relatively simple (a) inputs (overlap size measured as number of texels covered by two or more charts). Our outputs are overlap-free and our method is 60 times faster.}
\label{fig:uvpm_overlap_Sibenik}
\end{figure}

\textit{Parallel GPU Packing Methods.}   
UVPackMaster \cite{uvpackmaster} is a closed-source commercial GPU-based packer. While not interactive, it often improves both runtime and stretch versus Xatlas. Unfortunately, it is not robust, and fails to terminate on ~4\% of inputs in our dataset (including some of the largest ones). Worse, it introduces significant inter-chart overlaps on 7.5\% of inputs (Fig.~\ref{fig:uvpm_overlap_Sibenik}a). Finally, while its interface allows gutter specification, in our experiments their outputs did not satisfy gutter requirements on over 90\% of inputs. TABI robustly processes all inputs in our extensive dataset (Sec~\ref{sec:results}) and strictly satisfies gutter specifications. 

Several packing strategies achieve real-time performance by leveraging GPU parallelism. Some \cite{mueller2018sas, hladky2019tessellated, hladky2021snakebinning} make strict assumptions on the shape 
of the charts being packed (e.g. individual triangles or rectangular patches), making them unsuitable for our setting. Recent methods \cite{Neff2022MSA,vining2025fastatlas} support general chart shapes by operating on the bounding boxes of the input charts. \cite{Neff2022MSA} rounds boxes to power-of-two sizes and packs them using a superblock-based scheme. FastAtlas \cite{vining2025fastatlas} significantly outperforms \cite{Neff2022MSA} in terms of packing quality. It parallelizes the fold-and-push packing strategy of Chameleon \cite{igarashi2001adaptive} and adapts it to the GPU, with one key modification: a prefix sum-based relaxation of the previously sequential \textit{folding} step that significantly speeds up computation by allowing greater parallelism at the cost of increased downscaling (Fig.~\ref{fig:fa_igarashi}a).
While fast, the stretch introduced by FastAtlas is significantly worse than that of offline alternatives. 

TABI follows this overall fold-and-push framework \cite{igarashi2001adaptive,vining2025fastatlas} but makes major modifications that significantly reduce stretch (Fig~\ref{fig:teaser},~\ref{fig:impact}, Sec~\ref{sec:results}) while retaining interactive performance.

\textit{Interactive Packing Applications.}
Interactive packing methods are useful for many applications in digital content creation, real-time rendering, and games. These include surface painting \cite{igarashi2001adaptive,carrhart2004paintingdetail}, procedural texturing \cite{carrhart2002meshedatlases,debry2007player}, light mapping \cite{arvo1986backward,SourceEngine}, virtual texturing systems \cite{Barrett2008}, shadow mapping \cite{doghramachi2018tile}, and texture space shading (TSS) \cite{baker2012rock}. Both \cite{Neff2022MSA} and \cite{vining2025fastatlas} specifically target TSS as the application domain. For digital content creation applications, the main goal of interactive packing is to provide artists with immediate feedback. When this is not possible, it is often necessary to resort to offline packing at the cost of artist productivity \cite{sousa2025fastashell}.

\section{Fold-and-Push Framework} \label{sec:fold_push}

\begin{figure}
\includegraphics[width=\linewidth]{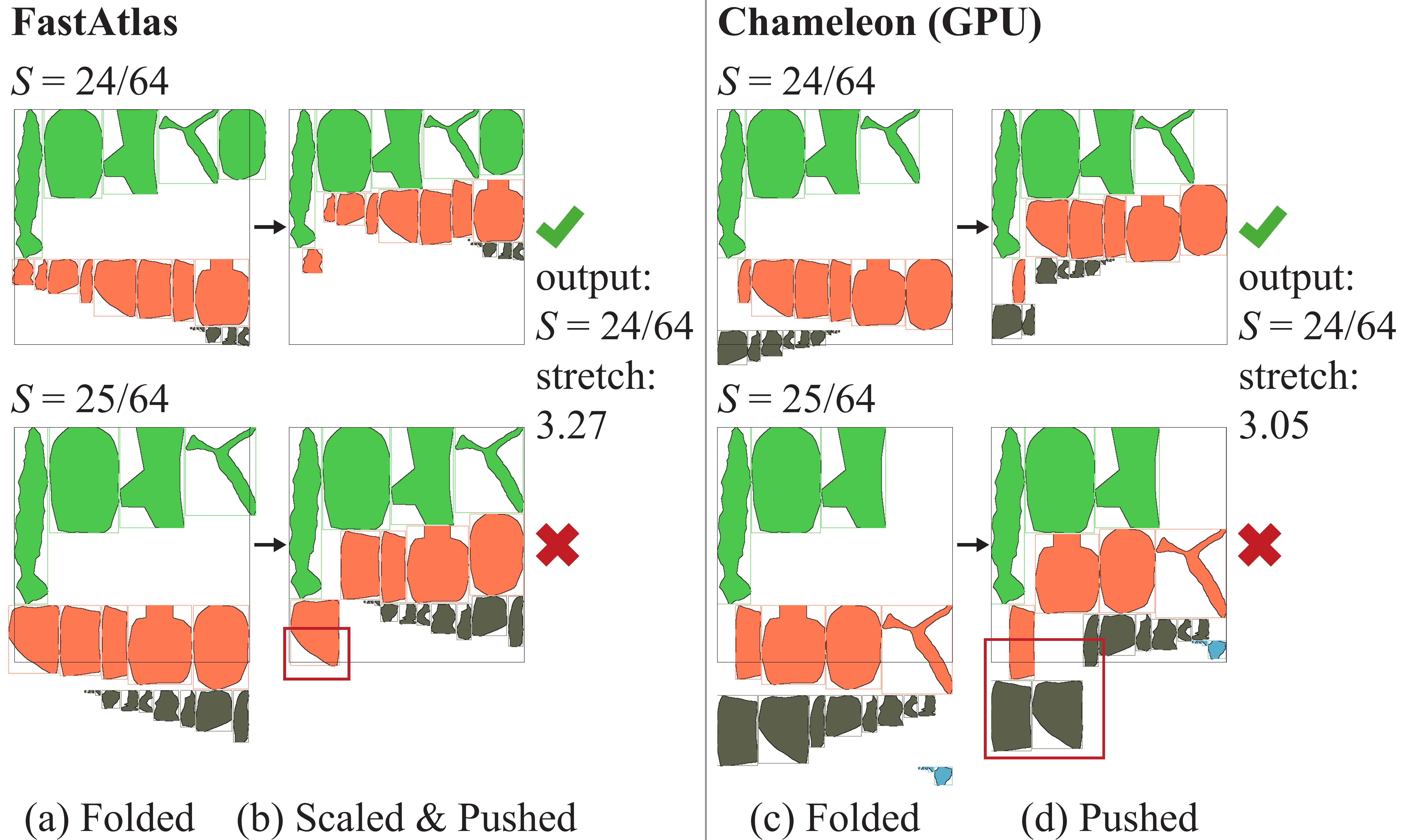}
\caption{
FastAtlas (left) and GPU Chameleon (right) outputs on the input shown in Fig. \ref{fig:overview}. Both methods operate on chart bounding boxes. FastAtlas folds rows using a prefix-sum based relaxation which causes them to initially overflow, and alternates one left-starting row with two right-starting rows (a); it must further downscale charts prior to pushing up in order to fit rows in the atlas (b). Chameleon folds rows sequentially and alternates between one left-starting row and one right-starting row (c), then pushes rows up (d).
Both methods generate packings with low tightness (note the large horizontal and vertical gaps between charts) and balance (note empty space in bottom right in Chameleon's output).
}
\label{fig:fa_igarashi}
\end{figure}

Achieving interactive packing performance on typical atlases requires a parallel, GPU based method. 
We therefore develop our TABI packing algorithm by extending the fold-and-push framework \cite{igarashi2001adaptive,vining2025fastatlas} (Fig~\ref{fig:fa_igarashi}). 
This parallelizable framework has been adapted to the GPU and outperforms prior parallel methods in terms of quality; it is thus a natural starting point for our method. 
Before proceeding to describe our method we briefly review packing methods that operate by folding and pushing charts (Fig. \ref{fig:fa_igarashi}, Sec~\ref{sec:background}) and analyse their shortcomings (Sec~\ref{sec:analyse}).

\subsection{Chameleon and FastAtlas}
 \label{sec:background}

The Chameleon packing algorithm operates on charts' bounding boxes. Boxes are rotated by $90^\circ$ if necessary so that they are taller than they are wide, and then sorted in decreasing height order to ensure that larger, and hence harder-to-pack charts, are packed first. Charts are packed into the atlas in two steps - {\em folding} and {\em pushing}. In the \textit{folding} step, sorted boxes are divided into rows which fit in the atlas width. Rows are then placed into the atlas from top to bottom, alternating between left-to-right and right-to-left directions, to try to prevent tall charts from bunching on one side of the atlas. In the \textit{pushing} step, the boxes in each row are pushed up as far as possible against previous rows.

Vining et al. \cite{vining2025fastatlas} adapt the packing strategy of Chameleon (Fig~\ref{fig:fa_igarashi}, right) to the GPU by parallelizing each step, and leveraging GPU parallelism to search for the optimal downscaling factor: folding and pushing are simultaneously performed over a range of candidate discrete scale factors $S$, and the largest $S$ where packing succeeds without overflowing the atlas is chosen. They then introduce FastAtlas, an even faster method (Fig~\ref{fig:fa_igarashi}, left). Unlike Chameleon, FastAtlas alternates between one left-to-right row and two right-to-left rows as a domain-specific heuristic targeting texture-space shading. FastAtlas parallelizes the folding step via approximation using a parallel prefix sum. After prefix-sum folding, the final chart in each row may overflow the atlas width, and all charts must undergo an intermediate scaling-down to bring rows back within the atlas width. This intermediate downscaling negatively impacts achieved packing scale, both on its own (Fig. \ref{fig:fa_igarashi}b) and because it requires FastAtlas to maintain additional horizontal space between charts compared to Chameleon in order to avoid gutter overlaps after downscaling.

\subsection{FastAtlas and Chameleon Challenges} 
\label{sec:explain} 
\label{sec:analysis}
\label{sec:analyse}

\begin{parWithWrapFigure}
\begin{wrapfigure}{l}{.27\columnwidth}%
\includegraphics[width=\linewidth]{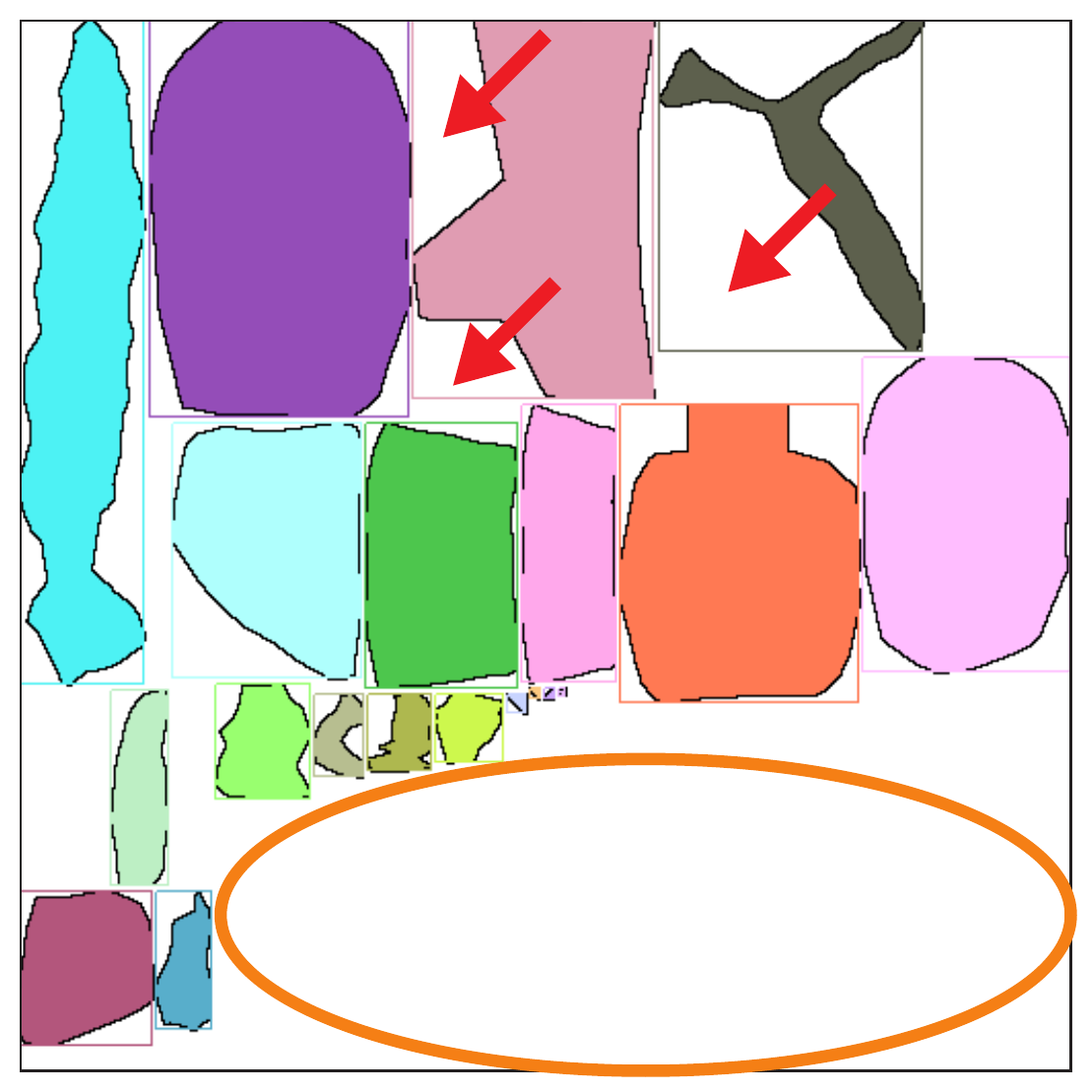}%
\end{wrapfigure}
While interactive, current fold-and-push methods produce notably lower quality atlases than Tetris-based approaches, suggesting room for improvement.  Analysing the outputs of fold-and-push approaches we observe that this sub-par performance is driven by two major sources of wasted space. First, since these methods are designed to operate on the bounding boxes of the charts, they waste the space between the charts and their bounding boxes (pointed to by arrows in inset); leading to undesirable horizontal and vertical gaps between packed charts. We refer to this property as lack of \textit{tightness}.
Second, packings generated by fold-and-push methods are often \textit{imbalanced}: charts on one side of the packing often reach the bottom of the atlas much earlier than those on the other, leaving a significant portion of the atlas unoccupied below the ``shorter'' side (highlighted by orange ellipse in inset). 
\end{parWithWrapFigure}

These observations motivate our packing strategy that reduces wasted space by improving both atlas tightness and balance while maintaining interactive performance.

\section{Method} \label{sec:algo} \label{sec:method}

Our method follows the general fold-and-push framework that balances quality against speed.  We improve atlas quality by addressing the tightness (Sec~\ref{sec:tight}) and imbalance (Sec~\ref{sec:balance}) challenges identified above. We use Chameleon-like row by row packing for smaller chart sets (Sec~\ref{sec:tight}-Sec.\ref{sec:summary}); and use a combination of row-by-row and prefix-sum based parallel packing for larger chart-sets (Sec~\ref{sec:optimizations}). This hybrid approach allows maintaining interactive performance across all inputs. 
Our input consists of a set of 2D charts to be packed, and the fixed dimensions of the atlas. We assume that charts are provided as polygonal meshes with coordinates defined in texel space (i.e. in the same units as the atlas dimensions). 
Our algorithm outputs per-chart scales and rigid transformations (translation, rotation, and/or reflection) that arrange the charts inside the atlas with no overlaps between the charts or their gutters. See supplementary for method pseudocode, parameter values, and details.

\begin{figure*}
\includegraphics[width=\linewidth]{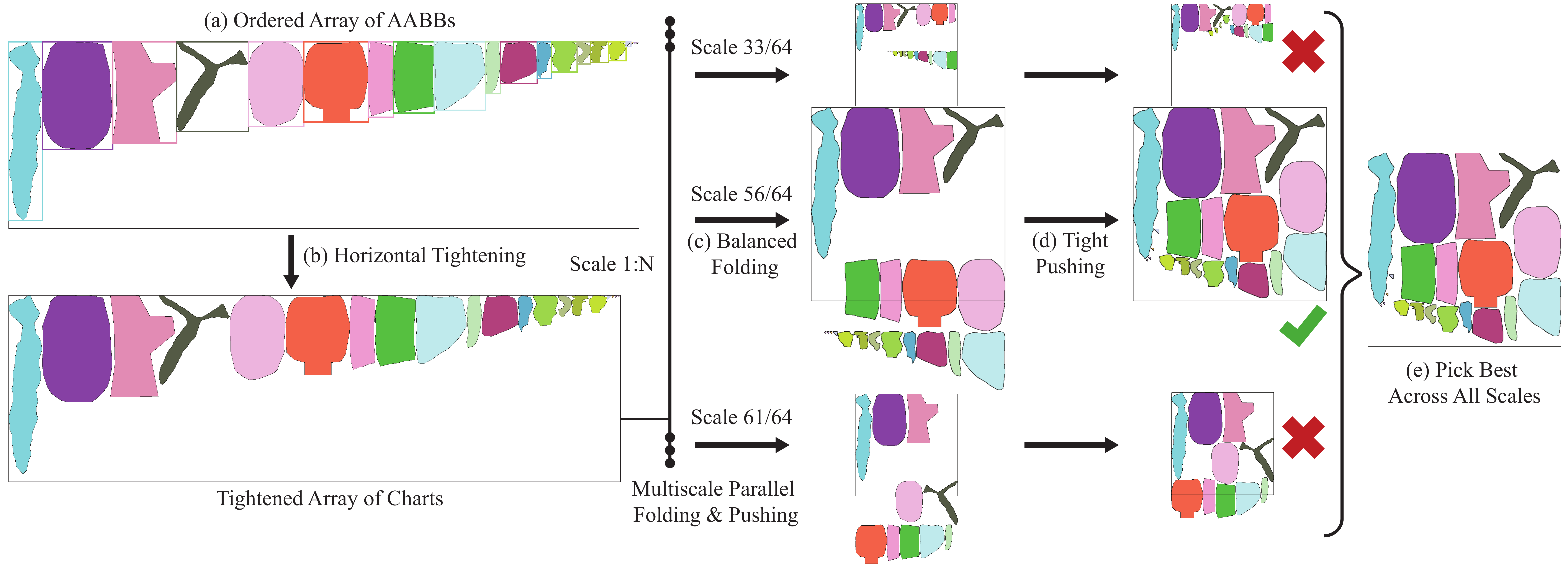}
\caption{TABI overview. (a) Charts placed in descending height order. They are then more {\em tightly} packed horizontally using per-chart proxy shapes (Sec~\ref{sec:tight_horizontal}). We compute a target chart scale by evaluating a discrete range of possibilities in parallel. For each candidate scale, charts are scaled and folded into rows using balanced folding (Sec.~\ref{sec:balance}). Charts are then tightly pushed up, one row at a time, using per-chart proxies (Sec.~\ref{sec:vertical_sliding}). Finally, we choose the largest scale that does not overflow the atlas.}
\label{fig:overview}
\end{figure*}

\subsection{Improving Packing Tightness} \label{sec:tight} 
\label{sec:tightness}
Improving packing tightness requires compacting horizontal gaps between charts during the folding step, and vertical gaps during the pushing step. 
Since the axis-aligned bounding boxes are already tightly packed, our tightening employs tighter chart bounding shapes  that can be computed and evaluated efficiently (Sec.~\ref{sec:proxies}).  We use these proxies to compact individual horizontal gaps in parallel for all charts (Fig~\ref{fig:overview}b) and use a similarly parallel approach to close vertical gaps during pushing (Fig~\ref{fig:overview}d). 

\subsubsection{Chart Proxies}
\label{sec:proxies}
 Evaluating gaps directly between polygonal charts is too expensive for our runtime budget. Raster-space computation (\`{a} la Tetris packing) would 
 require rasterizing and storing charts at every candidate scale factor which in turn would necessitate a  prohibitive amount of texture memory.
We avoid both pitfalls by by operating on tight bounding \textit{proxies} of chart shapes. We employ two such proxies in our method.

\begin{parWithWrapFigure}
\begin{wrapfigure}{l}{.285\columnwidth}%
\includegraphics[width=\linewidth]{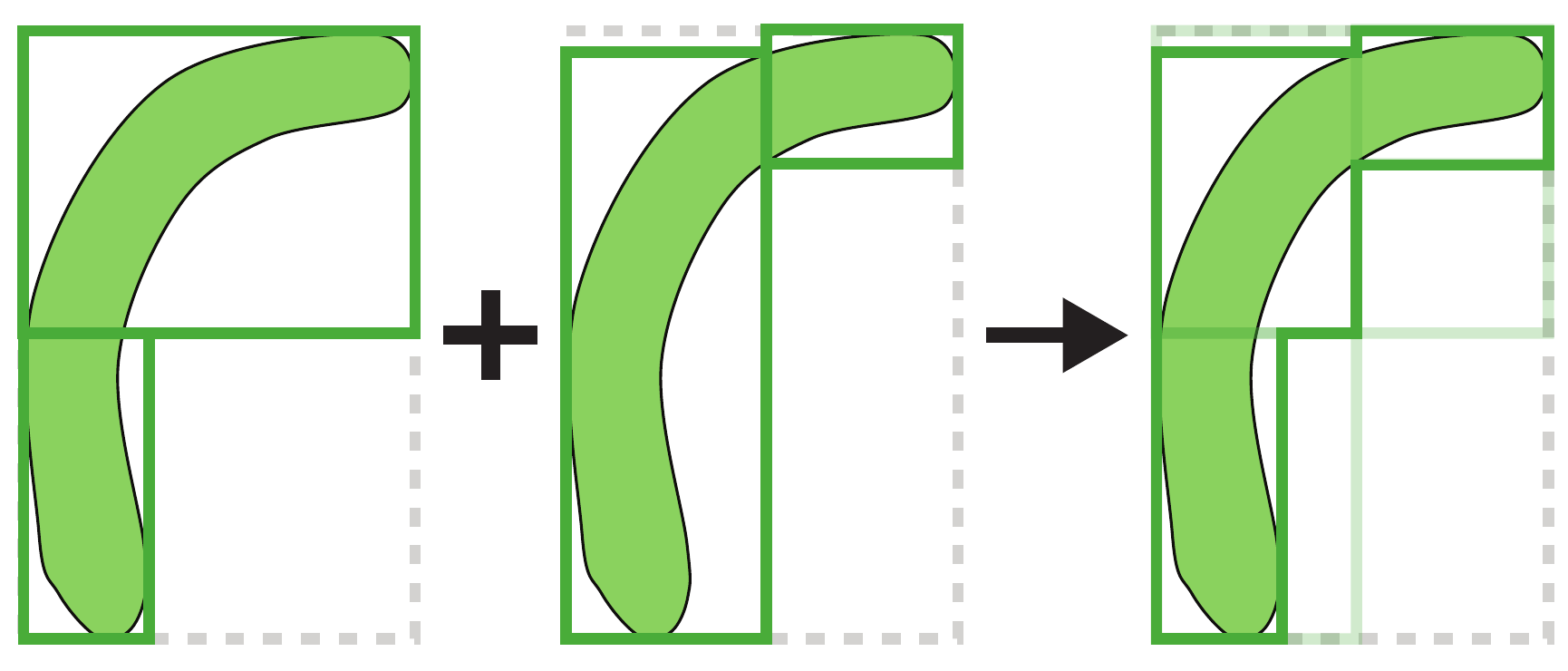}%
\end{wrapfigure}
Our first proxy is computed by dividing the chart into intervals of equal width along each of the x and y axes, then computing \textit{local AABBs} for the ``slice'' of the chart inside each interval; this can be efficiently done in parallel per-chart. The local AABBs along the x-axis effectively form a piecewise constant approximation of the chart's top and bottom boundaries, and those along the y-axis do the same for the left and right boundaries. After computing both sets of local AABBs, we then \textit{merge} them by taking the tighter bound along each axis to further refine our approximation of the four boundaries (see inset example, with 2 AABBs per axis).  The local AABB proxies capture concavities and void spaces \cite{limper2018boxcutter} in charts.
The more local AABBs are allocated, the better this approximation, enabling a tradeoff between quality and performance.
\end{parWithWrapFigure}

\begin{parWithWrapFigure}
\begin{wrapfigure}{l}{.13\columnwidth}%
\includegraphics[width=\linewidth]{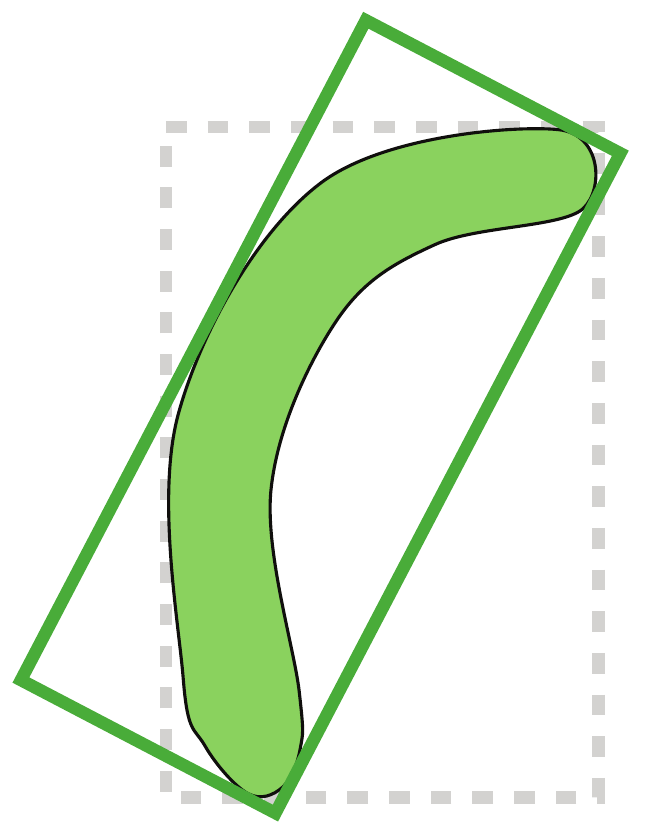}\\
\end{wrapfigure}
We augment the local AABBs by using a second proxy - the oriented bounding boxes (OBBs) for each charts (inset). While finding an exact OBB is too expensive for our interactive budget, we can obtain an approximate OBB by testing a fixed set of rotations in parallel and selecting the one that results in the minimal area box. 
\end{parWithWrapFigure}

\begin{parWithWrapFigure}
\textit{Optimizing Chart Orientations.}
The amount of horizontal and vertical compacting that can be achieved between two chart proxies depends on the relative orientation (i.e. horizontal or vertical reflection) of the underlying charts. 
\begin{wrapfigure}{l}{.55\columnwidth}%
\includegraphics[width=\linewidth]{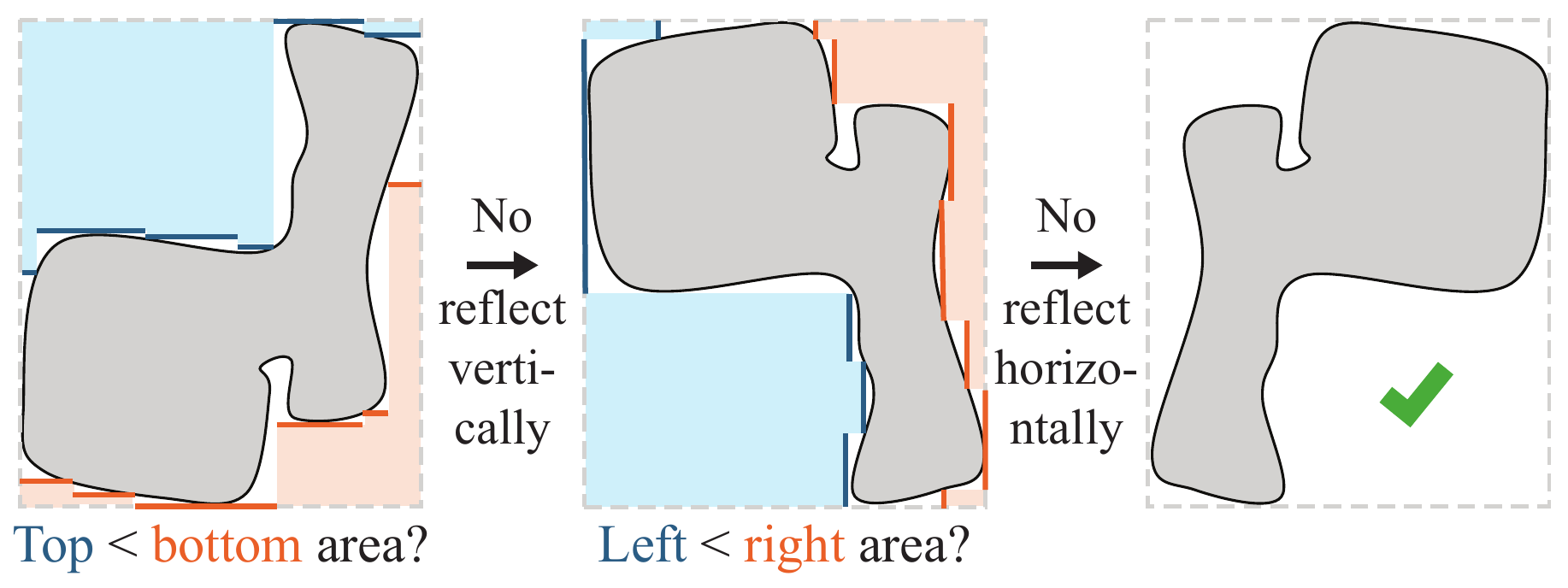}%
\end{wrapfigure} It would be prohibitively expensive to globally optimize the orientations of all pairs of charts, since each chart's choice of orientation affects both its neighbours. Instead, we independently compute favorable chart orientations, designed to allow for maximal compatibility between their proxies. 
Recalling that charts are lined up from left to right prior to folding the line, we aim to maximize empty area along the bottom of the chart, to maximize how far the next row can move up in the pushing step; and along the right of the chart, to maximize how far the next chart can move horizontally toward this one. We use our local AABB proxy, which captures the shape of charts' boundaries well, to determine whether reflection is necessary to satisfy these criteria.
\end{parWithWrapFigure}

\subsubsection{Compacting Horizontal Gaps}
\label{sec:tight_horizontal}

\begin{parWithWrapFigure}
The original folding step operates on a line of charts whose AABBs are side-by-side (aligned at the top; Fig. ~\ref{fig:overview}, top row). This line is guaranteed to be overlap-free, but is not tight. Starting from these initial positions, we can compact the horizontal gaps within the line by using our proxies and moving each pair of adjacent charts closer together by the maximum distance possible before they would intersect (Fig. ~\ref{fig:overview}b.). This \textit{compacting distance} can be calculated for each pair of charts in parallel. 
\begin{wrapfigure}{l}{.45\columnwidth}%
\includegraphics[width=\linewidth]{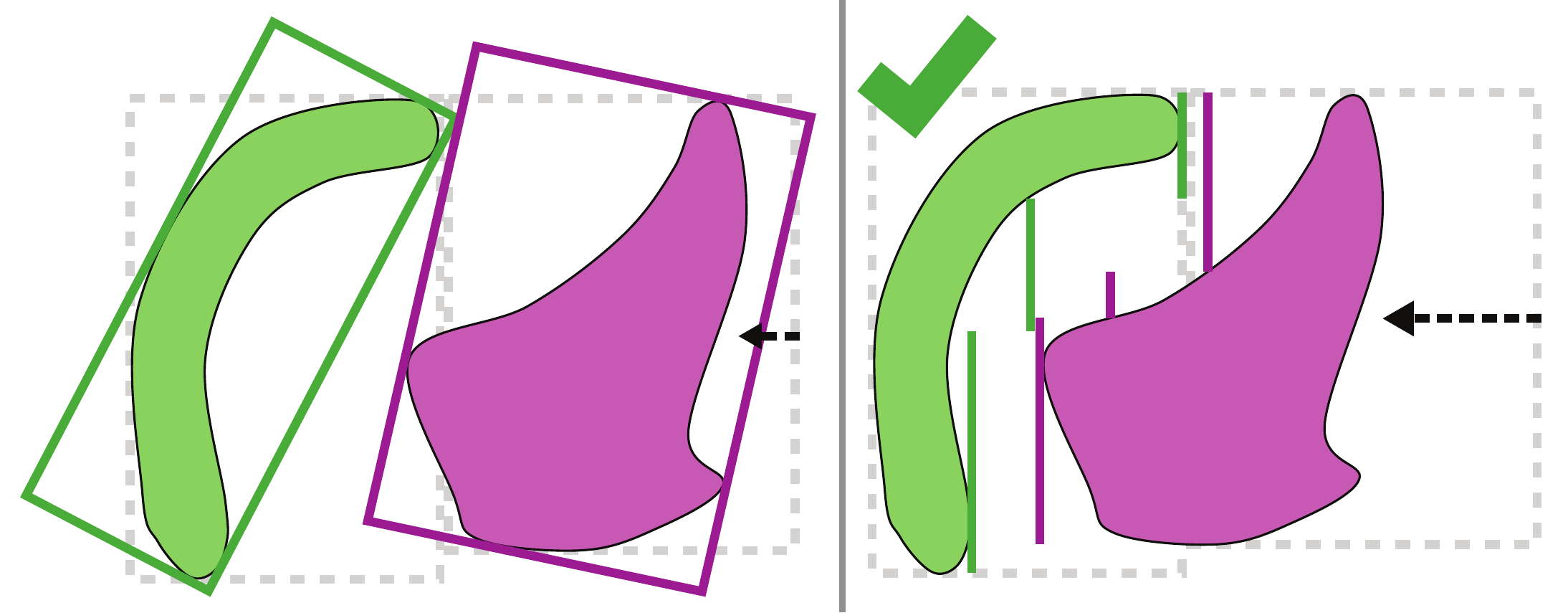}\\
\includegraphics[width=\linewidth]{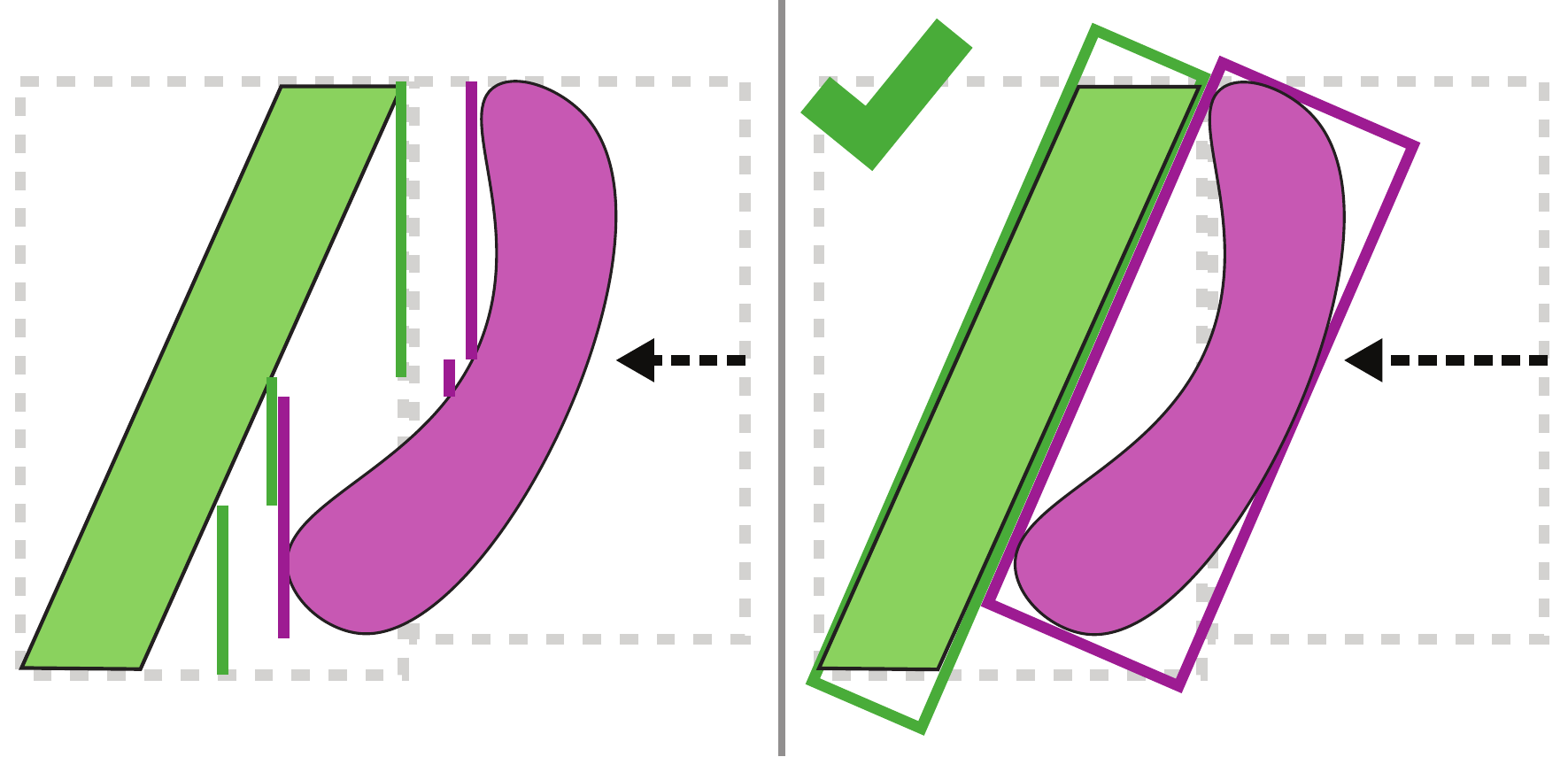}
\end{wrapfigure}
We use both proxies in tandem to achieve the best packing results by computing compacting distances from both proxies (indicated by arrows in the inset), and using the larger of the two compacting distances. To compute the compacting distance between a pair of local AABB proxies, we iterate over the first chart's right boundary and the second chart's left boundary and take the minimum distance between the piecewise constant values. 
We apply the Separating Axis Theorem \cite{schneider2002geometric} to find the horizontal distance between the OBBs.
\end{parWithWrapFigure}

\subsubsection{Compacting Vertical Gaps} 
\label{sec:vertical_sliding}

Our proxies also allow us to compact vertical gaps between rows during the pushing step (Fig. ~\ref{fig:overview}d). At this point, charts have known horizontal locations in the atlas, but their vertical locations are still unknown. Like FastAtlas, our parallelized pushing method operates using an advancing frontline -- an array storing an entry for each column of texels in the atlas, representing the packing height at that point. In FastAtlas, this frontline is updated one row at a time; each box in the first row writes its height to the frontline, and subsequent rows then set their vertical offset to the smallest value that prevents them from overwriting the frontline and write that value plus their height to the frontline texels they cover. We observe that this frontline does not require charts to be represented by AABBs. Accordingly, we can test whichever of the top boundaries of the proxies most tightly bounds the chart to find a tight vertical placement against the existing frontline. In practice, for each $x$ coordinate, this is whichever of the two proxies has the largest $y$ coordinate along its top edge. We then write the tightest fitting bottom boundary of the two proxies into the frontline after each update (whichever of the two proxies has the smallest $y$ coordinate for each $x$ coordinate on the bottom edge.)

\begin{parWithWrapFigure}
Pushing all charts up independently, however, can introduce inter-chart overlaps (inset, left). This is because the horizontal compacting distances we computed are only exact if chart pairs remain in their initial relative positions (i.e. are aligned at the top); if one chart moves up after horizontal compacting, they may now intersect (inset, left). 
\begin{wrapfigure}{l}{.5\columnwidth}%
\includegraphics[width=\linewidth]{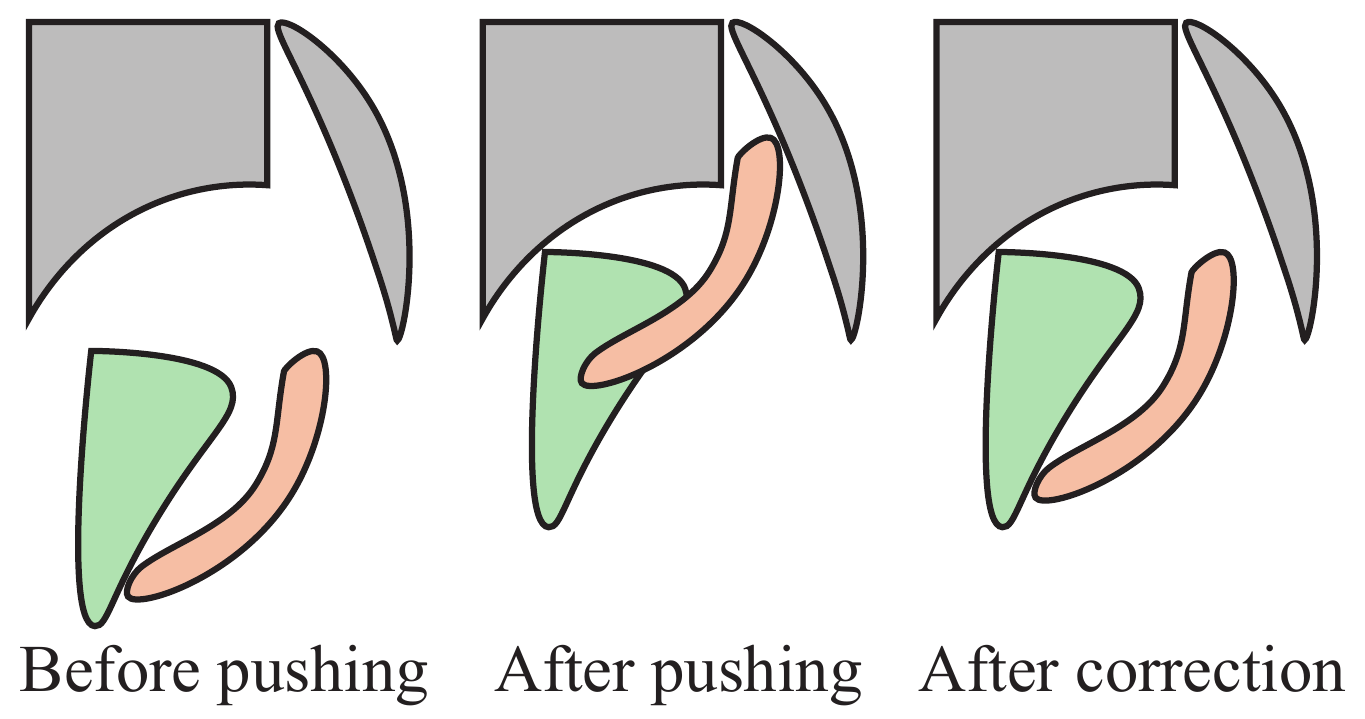}%
\end{wrapfigure}
In some cases, one chart in the pair can still move up freely without risk of intersecting the other; in other cases, neither chart can move up freely without potential intersection.  When compacting charts horizontally we identify all adjacent chart pairs where moving one chart up relative to the other can trigger intersection and record which chart's movement can trigger intersection. During the pushing stage we first allow all charts in a row to push all the way to the frontline. For each chart whose movement can trigger intersection, in parallel, we check if it moved up relative to the neighbor it can potentially intersect, and move it down to match the neighbor's vertical position. We repeat this process until all such chart pairs are vertically aligned.
\end{parWithWrapFigure}

\begin{figure}
\includegraphics[width=\linewidth]{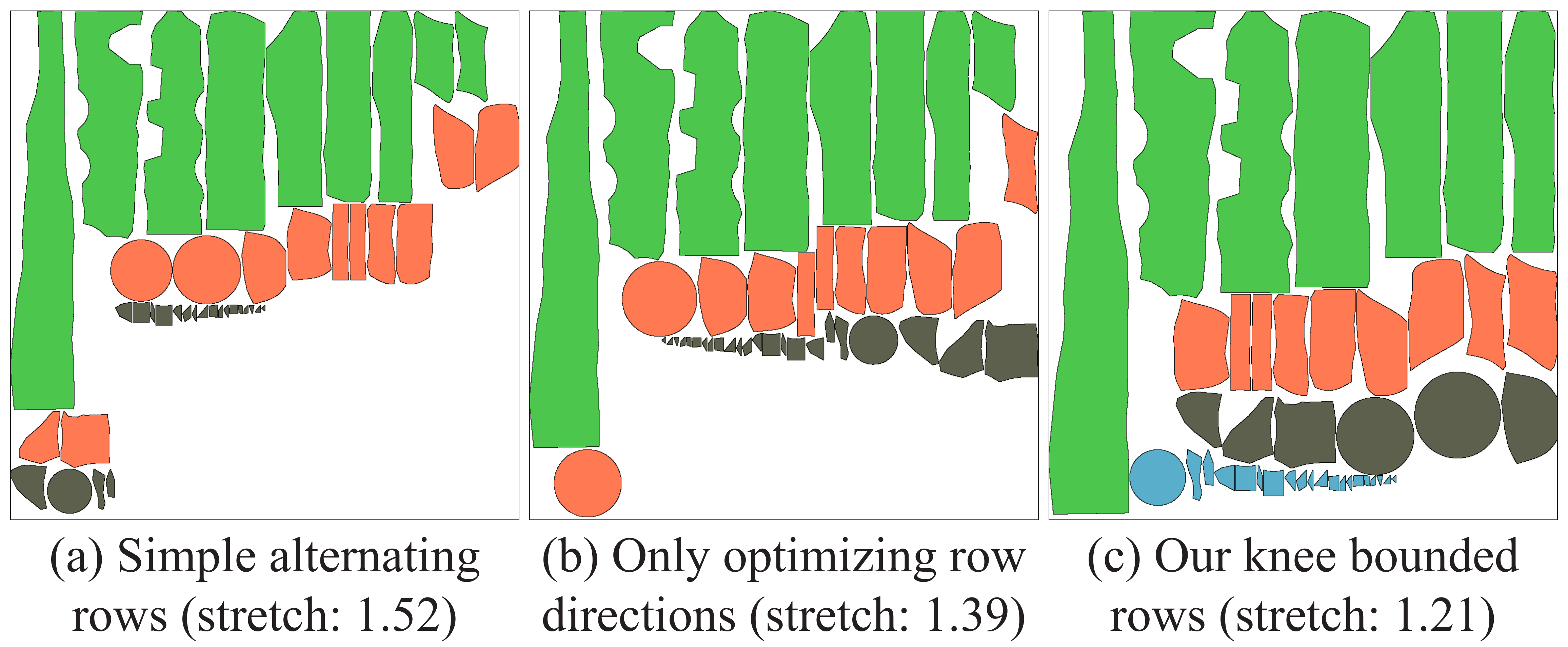}
\caption{Imbalanced packings are caused by a failure to equalize accumulated chart heights on the two sides of the packing. Simply alternating rows (left) immediately produces an unbalanced packing. Optimally selecting row directions to equalize height distribution (center) improves balance, but may still lead to wasted space. Our {\em knee-based} solution (right) handles large within-row changes in chart height, resulting in an efficient and balanced packing.}
\label{fig:knees}
\end{figure}

\subsection{Balanced Folding} 
\label{sec:balance} 
\label{sec:balanced}

We propose a novel folding method designed to avoid imbalance.  We observe that imbalance in the outputs of our baseline methods occurs due to their use of \textit{static} row alternation rules that  fail to equalize accumulated chart heights on the two sides of the packing (Fig. ~\ref{fig:knees}a).  We further observe that in many input chart sets, whether artist-created or algorithmically generated, the height of consecutive charts in our sorted line may differ dramatically (see first and second charts in Fig. ~\ref{fig:knees}a); such rows introduce significant imbalance between the accumulated heights on the left and right. 
Our first observation suggests that producing balanced packing, at a minimum, requires dynamically selecting whether chart rows are packed right-to-left or left-to-right at each row (Fig. ~\ref{fig:knees}b). However, when chart heights are highly non-uniform within a row, even optimal selection of row directions may be insufficient to balance the packing (Fig. ~\ref{fig:knees}b). Generating balanced packings therefore requires both a more robust strategy for selecting row directions, and a way to handle rows of highly non-uniform height, ones containing sharp jumps in chart height. We refer to these sharp jumps in chart height as \textit{knees} \cite{antunes2018knee}.%

Our key insight is that when these knees are present, folding the subsequent row(s) at the full atlas width independent of direction is sub-optimal as it will simply further increase the height on both sides of the packing, preserving the imbalance. Instead, we exploit the fact that knees create deep indentations, or concavities, in the frontline, ones that can be effectively filled by {\em shorter} rows.
Our method detects these knee concavities and folds subsequent rows at concavity width, effectively re-balancing the packing by filling the indentation with shorter rows, before continuing (Fig. ~\ref{fig:knees}c).

\textit{Knee Concavity Computation.}
We define a height difference between consecutive charts in a rows as a knee if this difference is sufficiently large in both absolute terms and relative to the height of the taller chart.
We use one knee per row to guide subsequent folding, selecting the one with the largest height difference if multiple exist.
The concavity-facing edge of the taller chart's AABB (right edge if row is left-to-right, left edge if row is right-to-left) provides a conservative estimate of the concavity's width. We refine this estimate by drawing a horizontal line at the height of the frontline texel adjacent to the current concavity edge, and updating the concavity edge to its closest intersection with the frontline. This process is implemented in parallel using atomic operations. 

\textit{Placing Rows.}
Since rows are pushed up sequentially, we delay the choice of a row's direction, folding width, and other per-row options until after all previous rows have been folded and pushed. This enables our method to read the current state of the packing and make the locally optimal decision, avoiding the pitfalls of prior static approaches. 

\begin{parWithWrapFigure}
\begin{wrapfigure}{l}{.45\columnwidth}%
\includegraphics[width=\linewidth]{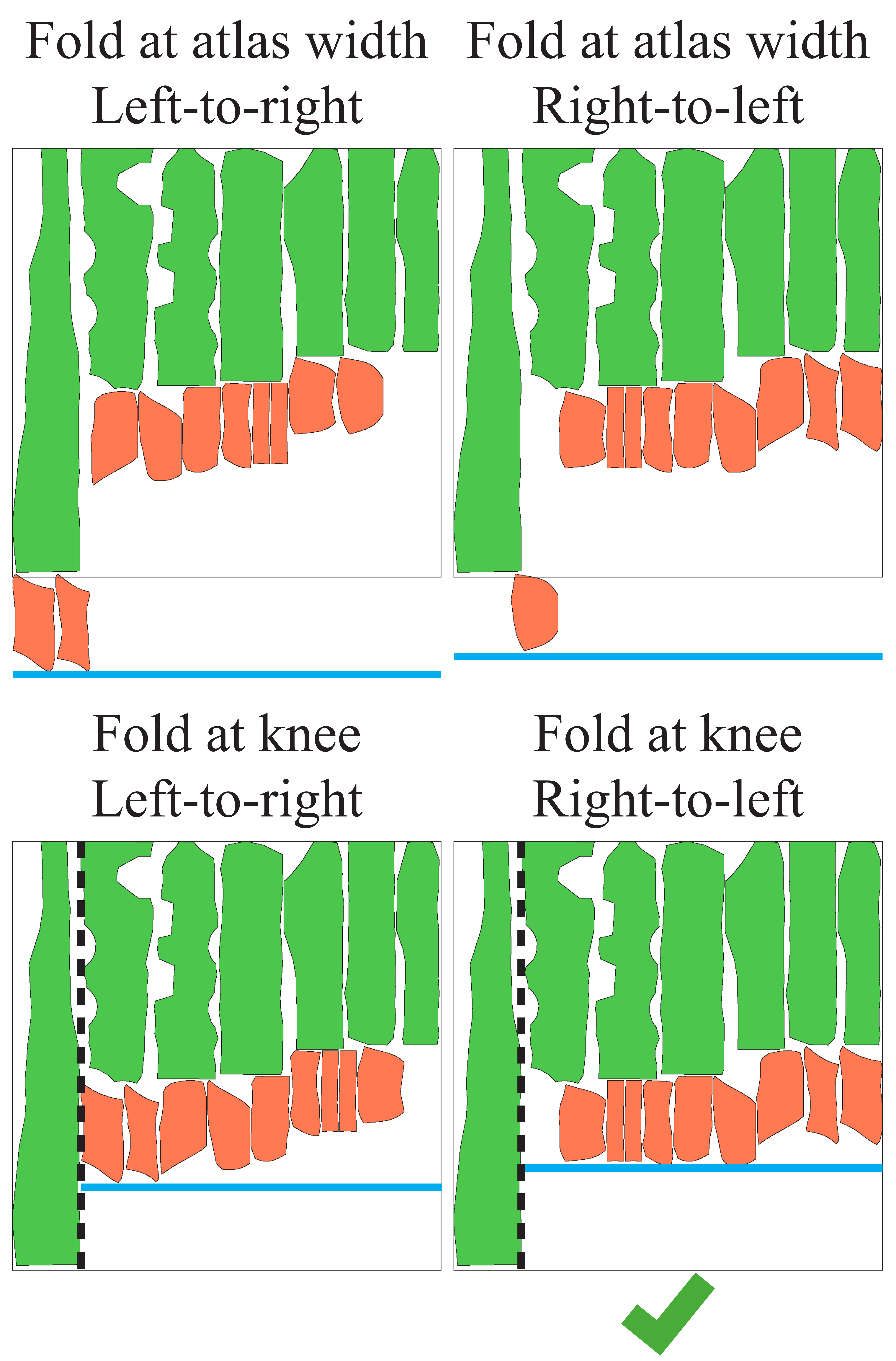}%
\end{wrapfigure}
We consider the following variables in placing each row. First, the row may be placed left-to-right or right-to-left. Second, the row may be folded at the width of the atlas or a knee concavity in a previous row (if one exists, and all intervening rows were also folded into the knee concavity). Lastly, we also consider folding each row with and without horizontal compacting. While the core benefit of horizontal compacting is its ability to fit more charts into each row, if it does not achieve this goal for a particular row, disabling it may be a better choice to enable charts to push up freely without interlock and leave more space between charts for future rows to push up into (below inset shows an example).
\end{parWithWrapFigure}

\begin{parWithWrapFigure}
\begin{wrapfigure}{l}{.35\columnwidth}%
\includegraphics[width=\linewidth]{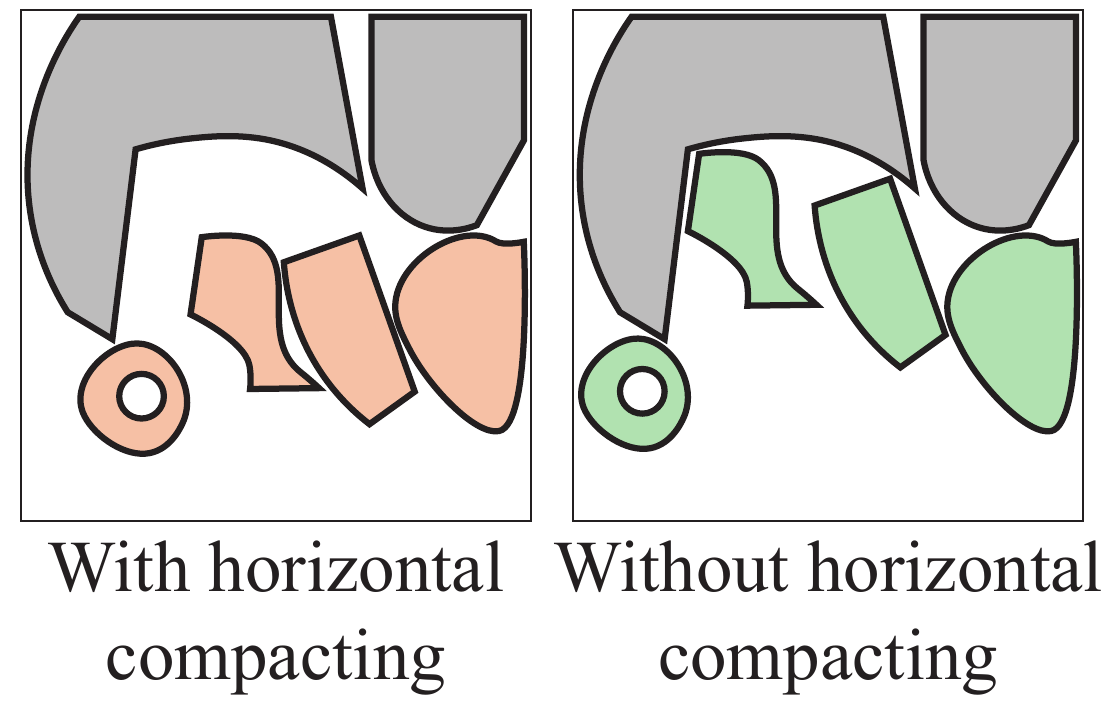}%
\end{wrapfigure}
In total, we consider 8 combinations of variables if a previous knee exists, and 4 otherwise. Due to charts' irregular shapes, it is challenging to predict which combination will yield the best outcome after folding and pushing. Instead, we directly evaluate folding and pushing for each potential combination.
\end{parWithWrapFigure}

We select the ``best'' combination through a hierarchical comparison process; each step compares pairs of combinations that differ in one variable, and keeps the better value. This allows us to apply different criteria at each hierarchy level. First, for each pair of combinations that differ by horizontal compacting, we enable horizontal compacting if it allows more charts to fit in the current row. We then consider row direction. We observe that packing fails when any point on the frontline exceeds the atlas height after pushing; in other words, the smaller the maximum height of the frontline, the further the packing is from failure. Thus, we select the row direction that minimizes the maximal frontline height (computed over the entire frontline for combinations folded at the atlas width, and over the knee concavity if folded at the knee, as the remaining texels are identical). Finally, if applicable, we determine whether to fold at the knee using the maximum height over the whole frontline. Intuitively, folding at the knee permits fewer charts to fit in the current row, but is desirable if it  improves packing balance. We therefore accept the knee-folded combination if its maximum height is at least marginally smaller than the alternative.

\subsection{Algorithm Summary} \label{sec:summary}
We can now summarize our TABI packing algorithm as follows. The initial steps of our algorithm follow Chameleon: we compute the AABBs of all charts, rotate the boxes to be taller than they are wide, and sort charts by AABB height. This is followed by two parallel passes which compute our shape approximations for each chart and determine each chart's orientation. We then compute horizontal compacting distances between each pair of adjacent charts in parallel. We simultaneously scale and pack charts at a range of discrete scale factors, and return the largest scale and packing that succeed. Packing proceeds row by row; each row evaluates folding and pushing for different placement configurations based on our knee folding strategy, and commits the best result based on the remaining height below the frontline. We implement our packing procedure as an OpenGL compute shader with one work group per scale factor in order to take advantage of barrier synchronization within a work group. The memory consumption of our method is linear in the number of charts; thus, given a maximum chart count, required memory can be determined in advance.

\subsection{Performance Optimization} 
\label{sec:optimizations}

Folding and pushing one row at a time is the primary runtime bottleneck of our method as input size increases. We observe that the initial rows, consisting of the largest charts, typically have the greatest impact on packing quality, and are likely to have the most knees. Furthermore, the amount of downscaling caused by using prefix-sum decreases as chart size gets smaller. Thus the benefits of sequential folding diminish once chart size becomes sufficiently small relative to atlas size. Therefore, for large chart sets (Fig~\ref{fig:optimization}), where by definition, many charts are tiny, 
the runtime cost of sequential folding outweighs its benefit in terms of quality.  We therefore transition between our high-quality sequential folding, and a modified version of  prefix sum folding which can fold all charts in a single parallel pass. 

We sequentially fold and push the first $N$ rows, dynamically selecting the best placement for each.  Then, we fold all remaining charts in parallel using an extension of the prefix sum approximation proposed by FastAtlas that incorporates our horizontal compacting. In FastAtlas, the prefix sum of the chart widths modulo the atlas width yields an approximate folding of charts into rows. We instead initialize element $i$ of the prefix sum array to the AABB width of chart $i$ minus the horizontal distance to chart $i + 1$; in other words, we compute the prefix sum of the horizontal \textit{offsets} between charts. After intermediate downscaling of the prefix sum-folded rows, we insert them into the atlas using the same fixed direction alternation as FastAtlas and push them up. Unlike FastAtlas, in this hybrid method, the final scale (product of $S$ and intermediate scale) may slightly differ between the first $N$ and the remaining rows. Thus, we return the candidate scale factor $S$ that maximizes the average final scale weighted by chart area without overflowing the atlas. We allow $N$ to vary per input by switching to prefix sum folding when no more knees are detected and the height of the tallest chart in the row decreases below a threshold $t_{opt}$, set to 1\% of atlas height.
Past this point, charts are small enough for intermediate downscaling to be negligible and the outcome of pushing is very similar regardless of row direction chosen. In practice, small charts often account for a significant fraction of large input chart sets, and our optimization enables us to process these charts much more quickly with negligible impact on packing scale.

\begin{figure}
\includegraphics[width=\linewidth]{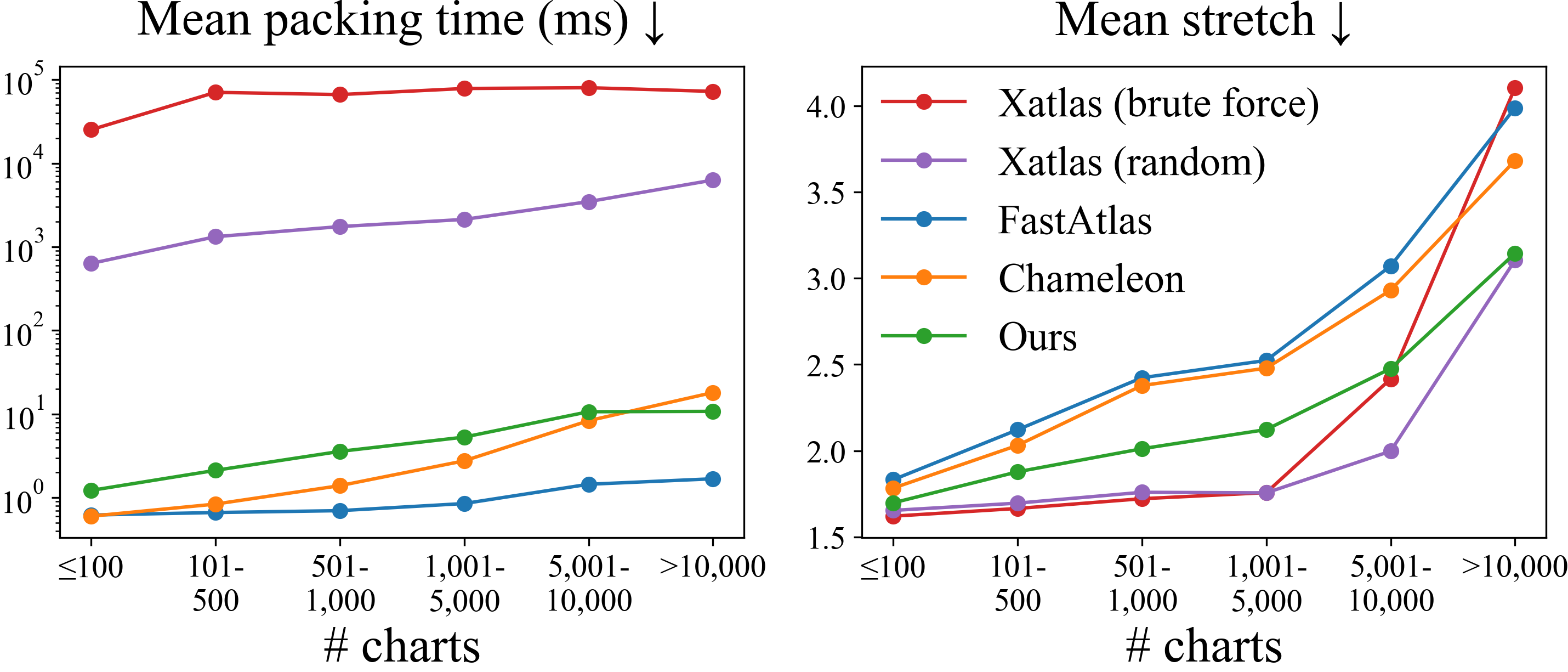}
\caption{Visual comparison of our method's runtime and stretch at different chart counts against alternatives (Xatlas \cite{xatlas}, FastAtlas \cite{vining2025fastatlas}, GPU Chameleon \cite{igarashi2001adaptive}) across our dataset.}
\label{fig:chart_count_vs_metrics}
\end{figure}

\begin{figure*}
\includegraphics[width=\linewidth]{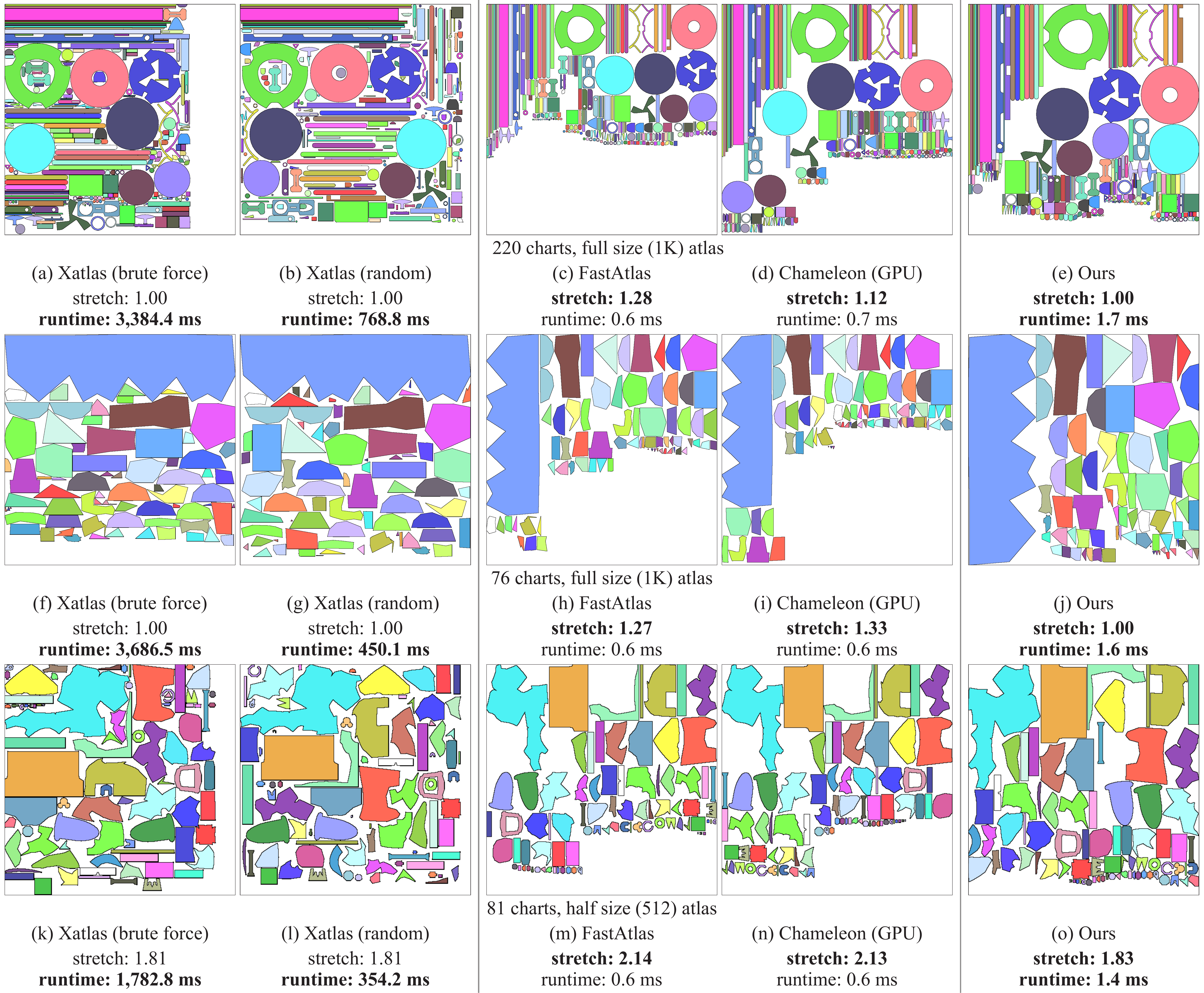}
\caption{Representative comparisons of our outputs against prior offline (Xatlas \cite{xatlas}) and interactive (FastAtlas \cite{vining2025fastatlas}, Chameleon \cite{igarashi2001adaptive}) alternatives on chart sets originating from UV unwrapping. We achieve $L^2$ stretch approaching that of offline methods, significantly outperforming FastAtlas and Chameleon, while maintaining interactive runtimes two orders of magnitude faster than the offline methods.
}
\label{fig:results_uv}
\end{figure*}

\begin{figure*}
\includegraphics[width=\linewidth]{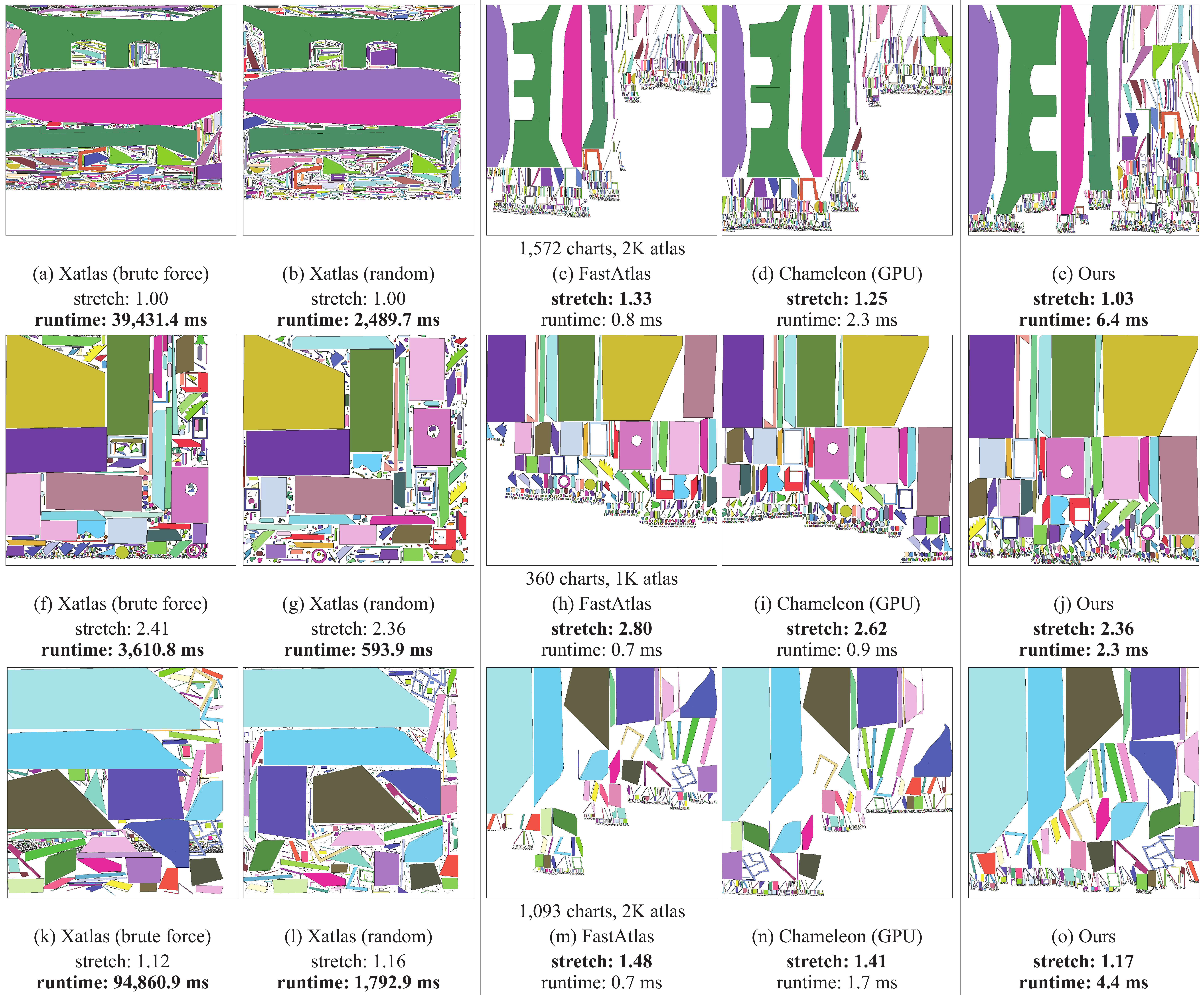}
\caption{Representative comparisons of our outputs against prior offline (Xatlas \cite{xatlas}) and interactive (FastAtlas \cite{vining2025fastatlas}, Chameleon \cite{igarashi2001adaptive}) alternatives on TSS sourced inputs. Our method achieves $L^2$ stretch approaching that of offline methods, outperforming FastAtlas and Chameleon, while maintaining interactive runtimes several orders of magnitude faster than those of offline alternatives.
}
\label{fig:results_tss}
\end{figure*}

\begin{figure}
\includegraphics[width=\linewidth]{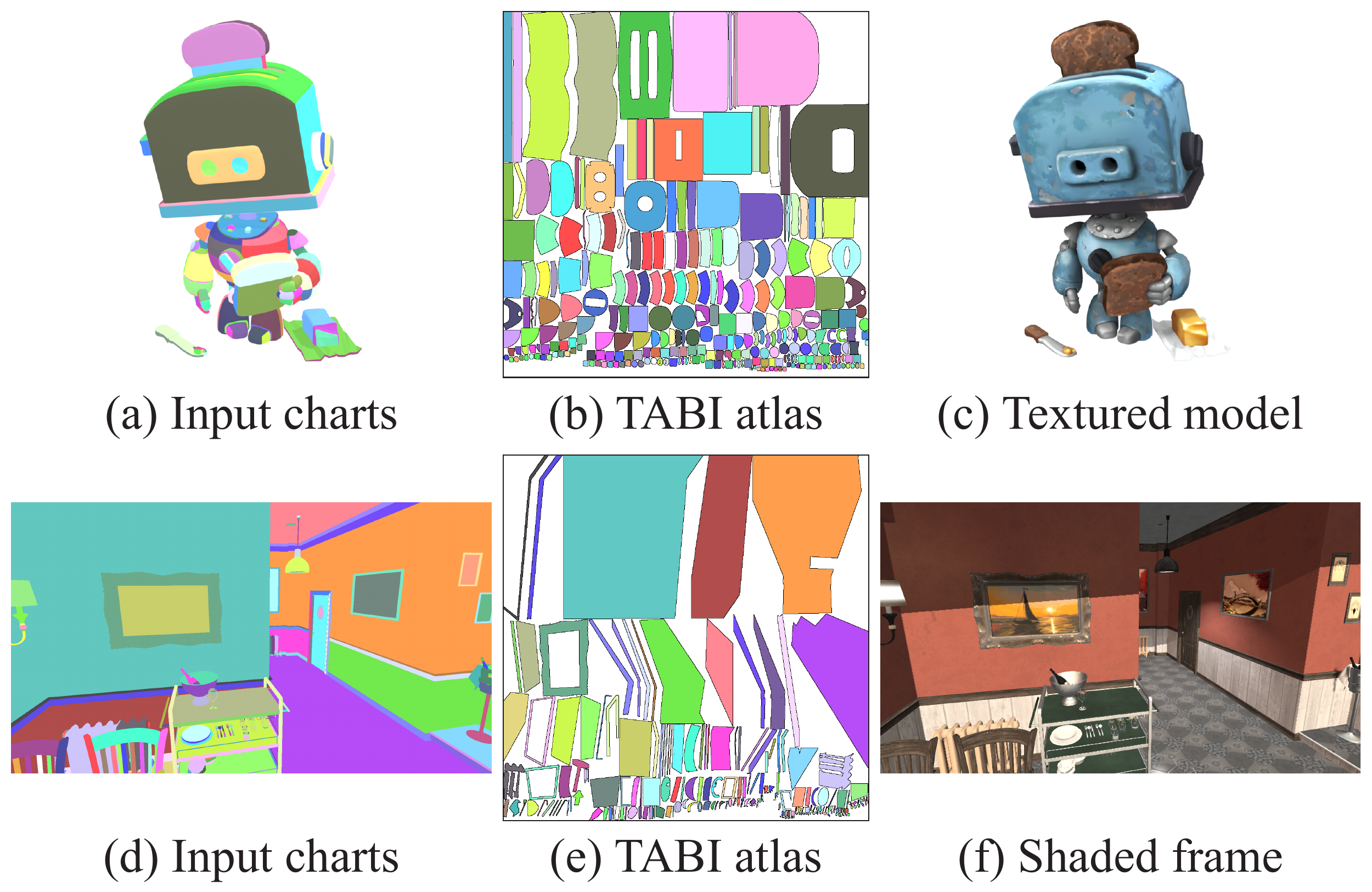}
\caption{Examples of chart sources in our dataset. Top row: (a) a 3D model with artist-created UV charts \cite{knodt2025texture}, (b) atlas packed by TABI, (c) textured 3D model using TABI packing. Bottom row: (d) a TSS frame charted by \cite{vining2025fastatlas}, (e) atlas packed by TABI, (f) shaded frame using TABI packing. ``Sad toaster'' model \textcopyright Tasha.Lime under CC BY-NC 4.0, via Sketchfab.}
\label{fig:results_3d}
\end{figure}

\section{Results} \label{sec:results}

\begin{figure}
\includegraphics[width=\linewidth]{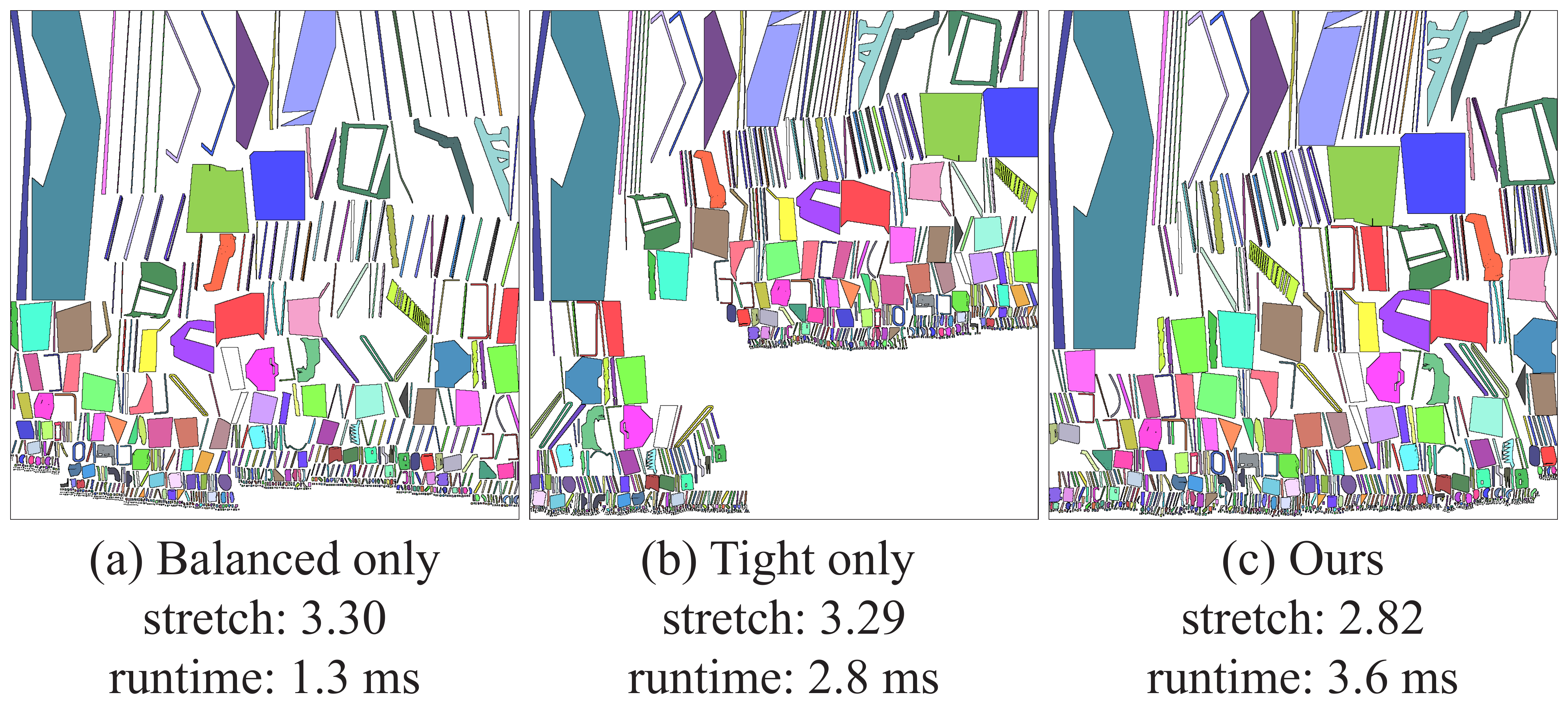}
\caption{Applying only balanced folding (Sec~\ref{sec:balanced}) (a) leaves empty space within charts' bounding boxes, while using only tightening (Sec~\ref{sec:tight}) (b) results in an imbalanced packing with significant unused space in the bottom right. Our full method (c) combines both components to achieve the best stretch.}
\label{fig:ablation}
\end{figure}

\begin{figure*}
\includegraphics[width=\linewidth]{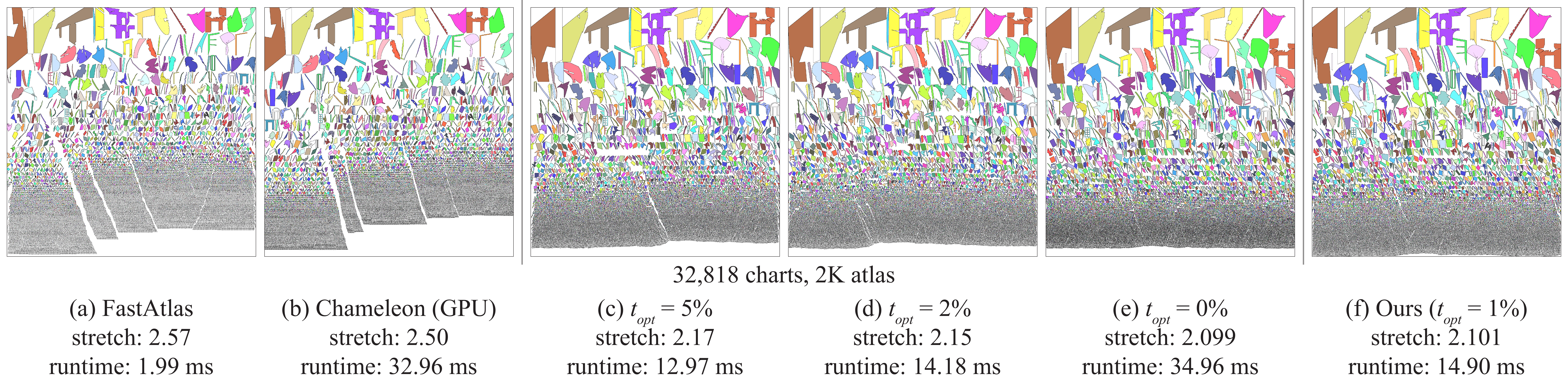}
\caption{Ablation of different values of $t_{opt}$ on an extremely large input. Higher values (c,d) provide greater performance improvements at the cost of higher stretch. A value of 0\% (e), i.e. no prefix sum folding, yields the best packing scale but requires a much longer runtime. Our chosen setting ($t_{opt} = 1\%$, f) significantly improves runtime with miniscule increase in stretch. At all values of $t_{opt}$ TABI produces better quality atlases than prior interactive alternatives (a, b).}
\label{fig:optimization}
\end{figure*}

We evaluate our method by packing 8,386 chart sets into square atlases of two different sizes each, for a total of 16,772 atlases. Our input chart sets include two different types of data, reflecting the typical settings where interactive atlassing would be used:
charts generated via UV unwrapping of 3D models, and texture-space shading (TSS) charts generated from camera paths through complex scenes \cite{vining2025fastatlas}. The former includes chart sets manually created by artists (63 from Sketchfab \cite{knodt2025texture}, 3 from TurboSquid plus the Utah teapot \cite{liu2017seamless}) as well as automatically unwrapped by software \cite{yang2023learningbased}.
Our chart sets range in size from 2 to 44,389 charts (30\% have under 1,000 charts, 50\% have 1,000 to 5,000 charts, 10\% have 5,000 to 10,000 charts, and 10\% have over 10,000 charts). Fig.~\ref{fig:results_3d} shows examples of source 3D models and TSS frames. Figs.~\ref{fig:teaser},~\ref{fig:overview},~\ref{fig:results_uv}-\ref{fig:optimization} show representative examples of atlases generated by our method and include atlases containing between 20 and 32,818 charts, Figs. ~\ref{fig:overview} and \ref{fig:optimization} respectively. Runtime and packing quality statistics are reported in Tab.~\ref{tab:runtime_all},~\ref{tab:stretch_all} and Fig.~\ref{fig:chart_count_vs_metrics}.

\begin{table}
\scriptsize
\setlength{\tabcolsep}{2pt}
\centering
\begin{tabular}{lc|cc|cc|c|ccc}
& & \multicolumn{5}{c|}{Mean packing time (ms) $\downarrow$} & \multicolumn{3}{c}{Packing time ratio} \\
                   &                    &                   & \textbf{Xatlas}    &                & \textbf{Cham-} &               & \textbf{Xatlas}   & \textbf{Fast-}   & \textbf{Cham-}   \\
                   &                    & \textbf{Xatlas}   & \textbf{(brute}    & \textbf{Fast-} & \textbf{eleon} &               & \textbf{(random)} & \textbf{Atlas}   & \textbf{eleon}   \\
\textbf{\# charts} & \textbf{\# inputs} & \textbf{(random)} & \textbf{force)}    & \textbf{Atlas} & \textbf{(GPU)} & \textbf{Ours} & \textbf{to ours}  & \textbf{to ours} & \textbf{to ours} \\
\hline
$\leq$ 100         & 1,726              & 633.51            & 25,189.01          & 0.62           & 0.60           & 1.22          & 492.70            & 0.54             & 0.52             \\
101-500            & 2,338              & 1,324.42          & 70,520.82          & 0.67           & 0.84           & 2.13          & 613.45            & 0.33             & 0.40             \\
501-1,000          & 1,640              & 1,749.25          & 66,288.21          & 0.70           & 1.39           & 3.58          & 475.74            & 0.20             & 0.39             \\
1,001-5,000        & 8,508              & 2,131.15          & 78,392.68          & 0.85           & 2.76           & 5.31          & 402.57            & 0.16             & 0.51             \\
5,001-10,000       & 814                & 3,475.95          & 80,130.26          & 1.45           & 8.34           & 10.68         & 323.28            & 0.14             & 0.77             \\
$>$ 10,000         & 1,746              & 6,298.64          & 72,350.89          & 1.69           & 18.12          & 10.79         & 589.70            & 0.16             & 1.66             \\
\textbf{All}       & \textbf{16,772}    & \textbf{2,326.34} & \textbf{70,091.95} & \textbf{0.90}  & \textbf{4.01}  & \textbf{5.11} & \textbf{464.03}   & \textbf{0.23}    & \textbf{0.62}   
\end{tabular}
\caption{Runtime comparison of our method against offline (Xatlas \cite{xatlas}) and interactive  (FastAtlas \cite{vining2025fastatlas}, GPU  Chameleon \cite{igarashi2001adaptive}) alternatives across our full dataset. The last 3 columns show the average ratios between alternatives and our runtimes. Our method is  two orders of magnitude faster than the fastest offline alternative (Xatlas random), and though slower than FastAtlas and Chameleon, it remains interactive.}
\label{tab:runtime_all}
\vspace{-3mm}
\end{table}

\begin{table}
\scriptsize
\setlength{\tabcolsep}{2pt}
\centering
\begin{tabular}{lc|cc|cc|c|cc}
& & \multicolumn{5}{c|}{Mean stretch $\downarrow$} & \multicolumn{2}{c}{Stretch improvement} \\
                   &                    &                   & \textbf{Xatlas} &                    & \textbf{Cham-} &               &                    & \textbf{vs.}   \\
                   &                    & \textbf{Xatlas}   & \textbf{(brute} &                    & \textbf{eleon} &               & \textbf{vs.}       & \textbf{Cham-} \\
\textbf{\# charts} & \textbf{\# inputs} & \textbf{(random)} & \textbf{force)} & \textbf{FastAtlas} & \textbf{(GPU)} & \textbf{Ours} & \textbf{FastAtlas} & \textbf{eleon} \\
\hline
$\leq$ 100         & 1,726              & 1.65              & 1.62            & 1.83               & 1.78           & 1.70          & 53\%               & 43\%           \\
101-500            & 2,338              & 1.70              & 1.67            & 2.12               & 2.03           & 1.88          & 53\%               & 41\%           \\
501-1,000          & 1,640              & 1.76              & 1.72            & 2.42               & 2.38           & 2.01          & 56\%               & 52\%           \\
1,001-5,000        & 8,508              & 1.76              & 1.76            & 2.52               & 2.48           & 2.12          & 50\%               & 48\%           \\
5,001-10,000       & 814                & 2.00              & 2.42            & 3.07               & 2.93           & 2.48          & 53\%               & 47\%           \\
$>$ 10,000         & 1,746              & 3.10              & 4.10            & 3.99               & 3.68           & 3.14          & 62\%               & 55\%           \\
\textbf{All}       & \textbf{16,772}    & \textbf{1.89}     & \textbf{2.00}   & \textbf{2.57}      & \textbf{2.48}  & \textbf{2.16} & \textbf{53\%}      & \textbf{48\%} 
\end{tabular}
\caption{Stretch comparison of our method against Xatlas \cite{xatlas} and interactive alternatives (FastAtlas \cite{vining2025fastatlas} and Chameleon \cite{igarashi2001adaptive}) across our dataset. Our outputs have significantly lower stretch than those produced by interactive alternatives. The last two columns show the amount by which our method closes the stretch gap between interactive alternatives and the best method per-input.}
\label{tab:stretch_all}
\vspace{-3mm}
\end{table}

\paragraph*{Experimental Setup.}
Our experimental settings reflect typical atlas generation scenarios in computer graphics where atlas sizes are square powers of 2 (due to support for mipmapping) and are similar to those of the state of the art methods in this space \cite{xatlas,vining2025fastatlas}.
We pack our UV-unwrapped charts (referred to as UV charts below) at two atlas sizes: the ``full'' size, which corresponds to the size of the original texture (manually packed by artists), and half of this full size. For the dataset of \cite{yang2023learningbased}, which does not have associated textures, we define the input chart size by scaling the original $[0,1]$ UV coordinates to a $1K \times 1K$ range, and use these dimensions as the ``full'' size; we then set half size to $512\times 512$.  The input size of our TSS charts is based on a $1920\times 1080$ screen resolution \cite{vining2025fastatlas}; we pack these charts at two different atlas sizes, $1K\times 1K$ and $2K \times 2K$. These sizes reflect real-life settings where downscaling may often be unavoidable, making packing non-trivial. 
We pre-rotate the UV charts to align their tight bounding boxes with the major axes prior to packing, a common optimization that improves packing quality. For TSS applications, rotating charts by angles which are not multiples of $90^\circ$ changes the sampling patterns of the chart content, which negatively impacts rendering quality and increases visual error; thus, we do not pre-rotate TSS data.
We configure TABI to leave 1-pixel gutters around charts to render borders for bilinear sampling, and no spacing along the 4 edges of the atlas, where ``clamped'' texture sampling avoids the need for gutters. To find the best downscaling factor, we follow FastAtlas and test 64 scales ranging from $\frac{1}{64}$ to $\frac{64}{64} = 1$ (see Figs ~\ref{fig:fa_igarashi},~\ref{fig:overview}). See supplementary for additional details.

\paragraph*{Robustness.} TABI successfully generated atlases on all the inputs tested and strictly satisfies the specified gutter sizes. As noted in Sec~\ref{sec:related}, and further detailed in the supplementary, this is not the case for alternatives such as UVPackMaster \cite{uvpackmaster}.  

\paragraph*{Measuring Downscaling.}
 Similar to \cite{vining2025fastatlas}, we evaluate the quality of generated packings by computing the $L^2$ stretch \cite{sander2001texture} between the triangles in the packed and input charts. The $L^2$ stretch corresponds to the average amount that the input signal is downsampled by packing, and predicts output visual quality. A stretch of 1 represents the ideal outcome when storing fixed-resolution surface signals, where all charts are packed successfully at exactly their input sizes; increased downscaling results in higher stretch. Tab~\ref{tab:stretch_all} reports the average stretch across all TABI outputs. 

\paragraph*{Runtime.} TABI takes 5.1 milliseconds on average to pack the inputs tested (Tab~\ref{tab:runtime_all}). Our average runtimes remain under 11 milliseconds even for atlases with over 10,000 charts, enabling interactive (60 fps) performance.  Only 1\% of our inputs require over 15 milliseconds to pack. Runtimes range from under a millisecond for smaller atlases, to 24 milliseconds on our most time consuming input.  
In addition to chart set size, the factors that impact packing time include the number of rows in the output atlas and the number of knees.  Roughly 60\% of the time is spent folding and pushing charts, 20\% is spent computing the tight approximations (Sec~\ref{sec:proxies}), and the rest is split roughly evenly between other parts of the method. All times computed on an AMD Ryzen 7 1800X CPU (32 GB  RAM) and an NVIDIA GeForce RTX 4090. 

\paragraph*{Comparisons.}
We compare our method both against existing packing methods capable of interactive performance, and against existing methods for offline packing. 
The latter can be seen as a reference baseline and a proxy for the best possibly achievable packing stretch.
We compare to FastAtlas \cite{vining2025fastatlas}, the current state-of-the-art in real-time atlassing. We also compare to our GPU implementation of Chameleon \cite{igarashi2001adaptive}, whose packing approach inspired the parallelized FastAtlas variation (Fig~\ref{fig:fa_igarashi}). We use Xatlas \cite{xatlas}, an industry-standard implementation of the Tetris packing algorithm (Section \ref{sec:related}) as our offline baseline. We include two variants of Xatlas; Xatlas ``random'' is the library's default setting and optimizes the search for candidate chart positions using randomization, while Xatlas ``brute force'' is intended to be slower but produce the best possible results (in our experiments the ``brute force'' method does indeed outperform the ``random'' on smaller chart sets, but seems to perform less well on larger ones). We use the same experiment settings described above for all methods we compare against (pre-rotate UV charts to align tight boxes with axes and use the same gutter settings as for TABI; allow all methods to rotate/reflect charts by multiples of $90^\circ$ only, etc). For a fair comparison, all methods search over the same set of 64 discrete scale factors to find the maximum scale at which packing succeeds. Since Xatlas does not provide built-in support for searching for the optimal downscaling factor, we follow prior work \cite{noll2011efficient} and run a binary search over the 64 scales. 

As Tab.~\ref{tab:runtime_all} shows, the Xatlas brute force method takes 43 seconds on average to atlas the inputs in our dataset; the most time consuming input takes it 32 minutes to atlas. Xatlas random takes 2 seconds on average to generate atlases across our input set, which is much too slow for the applications we target; TABI is almost 500 times faster. While slower than FastAtlas and Chameleon, it remains interactive and produces dramatically lower stretch packings across the inputs tested, narrowing the gap between interactive and offline methods by approximately 50\% on average. 

\paragraph*{Ablations.} We ablate our key algorithmic choices. Tab.~\ref{tab:ablationBalancedTight} and Fig~\ref{fig:ablation} validate our focus on improving tightness and balance as key factors in improving stretch compared to FastAtlas and Chameleon. Tightening the atlases via horizontal and vertical compacting reduces average stretch from 2.48 (Chameleon) to 2.33; while improving balance without tightening reduces it to 2.27. TABI combines both components and decreases the average to 2.16 (as a reference, the baseline Xatlas random average stretch is 1.89). This ablation suggests that the two components have roughly similar impact on the output quality.  

We ablate our runtime optimization strategy (Sec~\ref{sec:optimizations}) in Tab.~\ref{tab:optimization}, Fig.~\ref{fig:optimization}. We observed that for atlases with 1,000 or fewer charts, switching to the prefix sum folding strategy for small charts increases, rather than decreases, runtime (due to the overhead of dispatching the additional prefix-sum shaders). Our experiments thus focus on atlases with over 1,000 charts. For these atlases, using the prefix sum folding strategy in Sec~\ref{sec:optimizations} substantially decreases runtime, with larger values of $t_{opt}$ leading to larger speed-up. Notably, even with no runtime optimization, our method takes 8.12 milliseconds on average on large (>1,000 charts) atlases; though absent optimization performance is not interactive for atlases with over 10,000 charts, it remains two orders of magnitude faster than Xatlas. As expected, the earlier we switch from our method to prefix-sum folding, the more the stretch increases. Even with $t_{opt} = 5\%$, however, our stretch is significantly lower than that of FastAtlas and Chameleon. Notably, the stretch of our unoptimized method ($t_{opt} = 0\%$) is \emph{better} than that of the offline Xatlas method (3.01 vs 3.1) -- suggesting that for extra large inputs our method may be a better overall alternative (faster and better). These ablations validate our strategy of only using optimization for atlases with over 10,000 charts and using a small $t_{opt} = 1\%$ value as a way to achieve interactive performance with minimal increase in stretch. See supplementary for additional comparisons and ablations. 

\begin{table}
\scriptsize
\setlength{\tabcolsep}{2pt}
\centering
\begin{tabular}{lc|ccc|ccc}
& & \multicolumn{3}{c|}{Mean packing time (ms) $\downarrow$} & \multicolumn{3}{c}{Mean stretch $\downarrow$} \\
                   &                    & \textbf{Balanced} & \textbf{Tight} &               & \textbf{Balanced} & \textbf{Tight} &               \\
\textbf{\# charts} & \textbf{\# inputs} & \textbf{only}     & \textbf{only}  & \textbf{Ours} & \textbf{only}     & \textbf{only}  & \textbf{Ours} \\
\hline
$\leq$ 100         & 1,726              & 0.66              & 0.96           & 1.22          & 1.74              & 1.72           & 1.70          \\
101-500            & 2,338              & 0.92              & 1.55           & 2.13          & 1.97              & 1.92           & 1.88          \\
501-1,000          & 1,640              & 1.26              & 2.59           & 3.58          & 2.19              & 2.17           & 2.01          \\
1,001-5,000        & 8,508              & 2.05              & 3.96           & 5.31          & 2.31              & 2.24           & 2.12          \\
5,001-10,000       & 814                & 5.02              & 7.93           & 10.68         & 2.70              & 2.63           & 2.48          \\
$>$ 10,000         & 1,746              & 3.48              & 8.61           & 10.79         & 3.46              & 3.33           & 3.14          \\
\textbf{All}       & \textbf{16,772}    & \textbf{1.97}     & \textbf{3.86}  & \textbf{5.11} & \textbf{2.33}     & \textbf{2.27}  & \textbf{2.16}
\end{tabular}
\caption{Ablating our full method against using only balanced folding or only tightening across our dataset (left: average packing time in milliseconds, right: average $L^2$ stretch). By optimizing  both tightness and balance we achieve the best stretch.}
\label{tab:ablationBalancedTight}
\vspace{-3mm}
\end{table}

\section{Conclusions}
\label{sec:conclusions}

We introduce TABI, a new interactive texture atlas packing method that significantly improves on the performance of previous work. Our tight and balanced packings significantly reduce downscaling by decreasing wasted space around each chart and by better balancing chart placement within the atlas. Compared to previous interactive methods, TABI reduces the gap in atlas stretch compared to offline atlases by 48\% on average, making it suitable for a wide range of applications where packing quality is critical while maintaining interactive performance.

\begin{table}
\scriptsize
\setlength{\tabcolsep}{2pt}
\centering
\begin{tabular}{lc|cccc|c|cccc|c}
& & \multicolumn{5}{c|}{Mean packing time (ms) $\downarrow$} & \multicolumn{5}{c}{Mean stretch $\downarrow$} \\
                   &                    & \textbf{$\mathbf{t_{opt}}$} & \textbf{$\mathbf{t_{opt}}$} & \textbf{$\mathbf{t_{opt}}$} & \textbf{$\mathbf{t_{opt}}$} &               & \textbf{$\mathbf{t_{opt}}$} & \textbf{$\mathbf{t_{opt}}$} & \textbf{$\mathbf{t_{opt}}$} & \textbf{$\mathbf{t_{opt}}$} &               \\
\textbf{\# charts} & \textbf{\# inputs} & \textbf{= 5\%}              & \textbf{= 2\%}              & \textbf{= 1\%}              & \textbf{= 0\%}              & \textbf{Ours} & \textbf{= 5\%}              & \textbf{= 2\%}              & \textbf{= 1\%}              & \textbf{= 0\%}              & \textbf{Ours} \\
\hline
1,001-5,000        & 8,508              & 4.62                        & 4.62                        & 4.70                        & 5.31                        & 5.31          & 2.14                        & 2.13                        & 2.13                        & 2.12                        & 2.12          \\
5,001-10,000       & 814                & 6.28                        & 6.41                        & 6.92                        & 10.68                       & 10.68         & 2.54                        & 2.52                        & 2.51                        & 2.48                        & 2.48          \\
$>$ 10,000         & 1,746              & 9.31                        & 9.72                        & 10.79                       & 20.60                       & 10.79         & 3.20                        & 3.17                        & 3.14                        & 3.01                        & 3.14          \\
\textbf{All}       & \textbf{11,068}    & \textbf{5.48}               & \textbf{5.56}               & \textbf{5.82}               & \textbf{8.12}               & \textbf{6.57} & \textbf{2.34}               & \textbf{2.33}               & \textbf{2.32}               & \textbf{2.29}               & \textbf{2.31}
\end{tabular}
\caption{ Ablating $t_{opt}$ on inputs with 1,000 charts or more (left: average packing time in milliseconds, right: average $L^2$ stretch between packed and input charts). Increasing $t_{opt}$ reduces packing time, but increases stretch. Our method ($t_{opt}=0\%$ for inputs with up to 10,000 charts and $t_{opt}=1\%$ otherwise), minimally increases stretch compared to $t_{opt}=0\%$ while achieving interactive runtimes.}
\label{tab:optimization}
\vspace{-3mm}
\end{table}

\begin{figure}
\includegraphics[width=\linewidth]{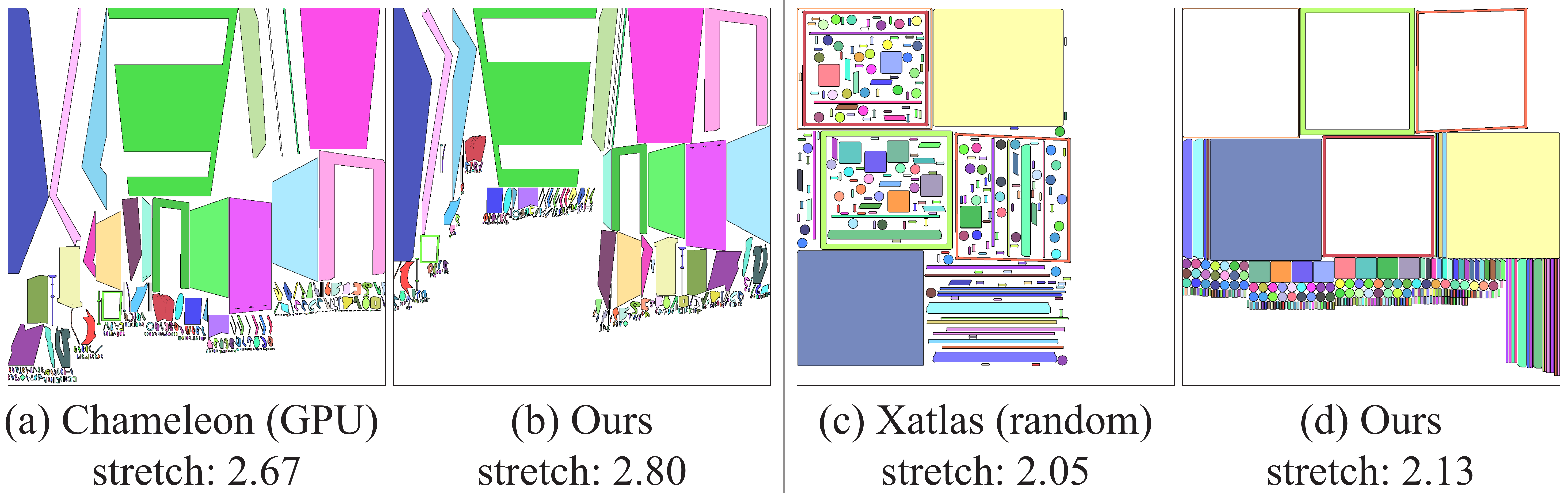}
\caption{(a,b) Chameleon (a) and our TABI (b) packings on the input where TABI performs worst relative to Chameleon across our dataset (even in this worst case, the difference in stretch is minor). (c,d) A representative example of inputs where Tetris type methods (c) outperform TABI (d) by placing small charts into interior voids.} 
\label{fig:limitations}
\end{figure}

\textit{Limitations and Future Work.}
TABI produces higher stretch atlases than Chameleon on only 2\% of inputs (and higher stretch atlases than FastAtlas on only 0.5\% of inputs). Most of these are due to the inherently heuristic nature of our method; worst example shown in Fig~\ref{fig:limitations}ab.  TABI performs worst compared to offline methods on inputs where even the tight approximations we use only loosely approximate the chart shape, see e.g. Fig~\ref{fig:limitations}cd. An interesting area of future work is to find a way to pack charts into interior voids at interactive speed.

\bibliographystyle{eg-alpha-doi} 
\newcommand{\etalchar}[1]{$^{#1}$}

\appendix

\section{Implementation Details}

We now describe additional details of our implementation of the TABI packing algorithm.

\paragraph*{Computing Local Axis-Aligned Bounding Boxes.} 
\label{sec:local_aabb_details}
Our local AABB proxy can be straightforwardly constructed in parallel. Each vertex and each edge (in case edges cross intervals) independently assigns itself to one or more intervals along each of the x and y axes, and updates the corresponding bounding boxes using atomic operations. Charts can then independently merge their two sets of local AABBs to obtain the final refined proxy.

\paragraph*{Computing Oriented Bounding Boxes.} 
\label{sec:obb_details}
To construct our OBB proxy, we test a fixed set of rotations and select the one that results in the minimal area box. We test 8 evenly spaced rotations in the interval $[0, \frac{7\pi}{16}]$, which we find provides a sufficient balance between accuracy and speed; testing rotations beyond $\frac{\pi}{2}$ is unnecessary because charts are rotated to be at least as tall as they are wide.

\paragraph*{Computing Chart Orientations.} 
\label{sec:orientation_details}
We describe how we leverage our local AABB proxy to orient charts in a way that maximizes the empty area to the bottom and right of the chart. We observe that the local AABB proxy provides piecewise constant approximations of the top, bottom, left, and right boundaries of each chart. Thus, the area between these boundaries and the edges of the chart's enclosing AABB approximates the empty area around the chart, which we seek to move to the bottom right through horizontal and/or vertical reflection. We first compute the empty area between the chart's top boundary and the top edge of its enclosing AABB (``top area'') and the corresponding ``bottom area''. If the top area is greater than the bottom area, we reflect the chart vertically to move the greater empty area to the bottom. Then, we compute the ``left area'' and ``right area'' similarly.
\begin{parWithWrapFigure}
\begin{wrapfigure}{l}{.425\columnwidth}%
\includegraphics[width=\linewidth]{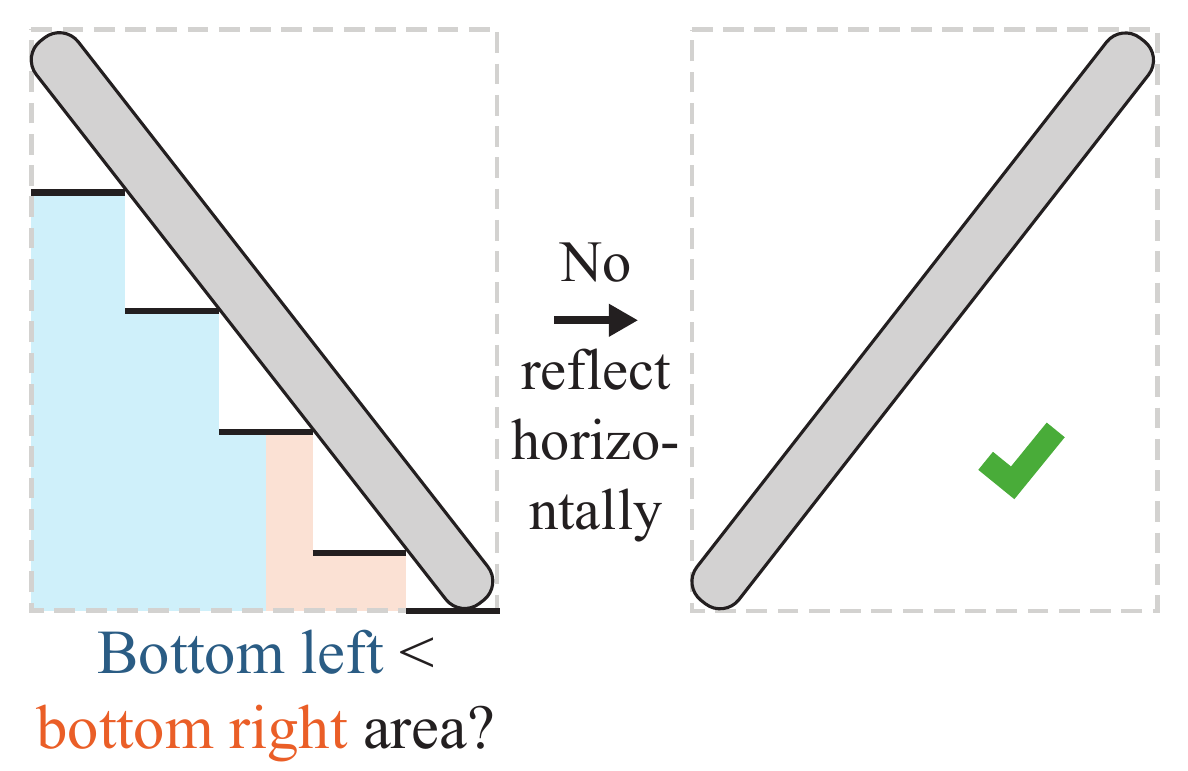}%
\end{wrapfigure}
If the left area is significantly greater than the right area, which we define as a difference greater than 10\% of the chart's AABB area between the left area and right area, we reflect the chart horizontally to move the greater empty area to the right. However, if the left and right areas are not significantly different (e.g. if the chart is perfectly diagonal, such as inset example), we try to maximize the empty area in the bottom right corner by comparing the portion of the bottom area located to the left of the chart's midline (``bottom left area'', blue area in inset) against the area to the right (``bottom right area'', orange area in inset). If the bottom left area is greater than the bottom right area, we also reflect the chart horizontally.
\end{parWithWrapFigure}

\paragraph*{Detecting Interlocking Charts.} 
\label{sec:interlock_details}
In order to eliminate potential overlaps during pushing, we mark in the horizontal compacting step which (if any) charts in each pair may trigger intersection if moved up relative to the other. If the horizontal compacting distance is zero, neither chart is restricted from moving up. Otherwise, it depends on which proxy provided the final compacting distance and thus the final non-overlapping positions.

\begin{parWithWrapFigure}
\begin{wrapfigure}{l}{.5\columnwidth}%
\includegraphics[width=\linewidth]{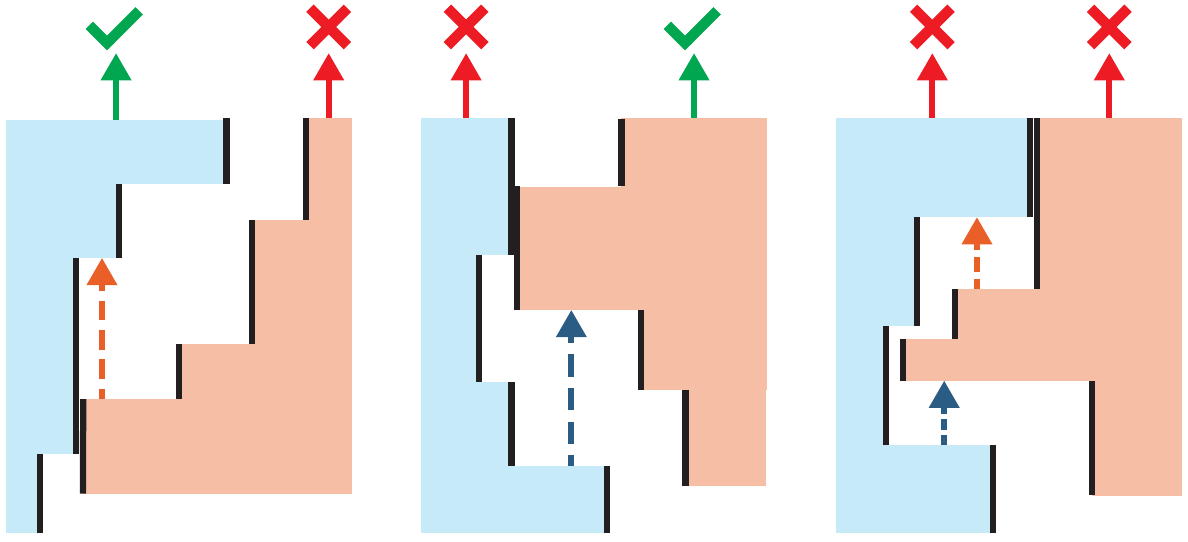}%
\end{wrapfigure}
If the local AABB proxy was used, one or both charts can potentially trigger intersection if moved up (red arrows with X in inset). We can determine if this is the case independently for each chart: a chart can move up without potential intersection (green arrows with checkmark in inset) if no rectangular segment formed by its piecewise constant boundary is ``below'' a segment of the other chart's boundary (violations indicated by dashed vertical arrows in inset).
\end{parWithWrapFigure}

\begin{parWithWrapFigure}
\begin{wrapfigure}{l}{.4\columnwidth}%
\includegraphics[width=\linewidth]{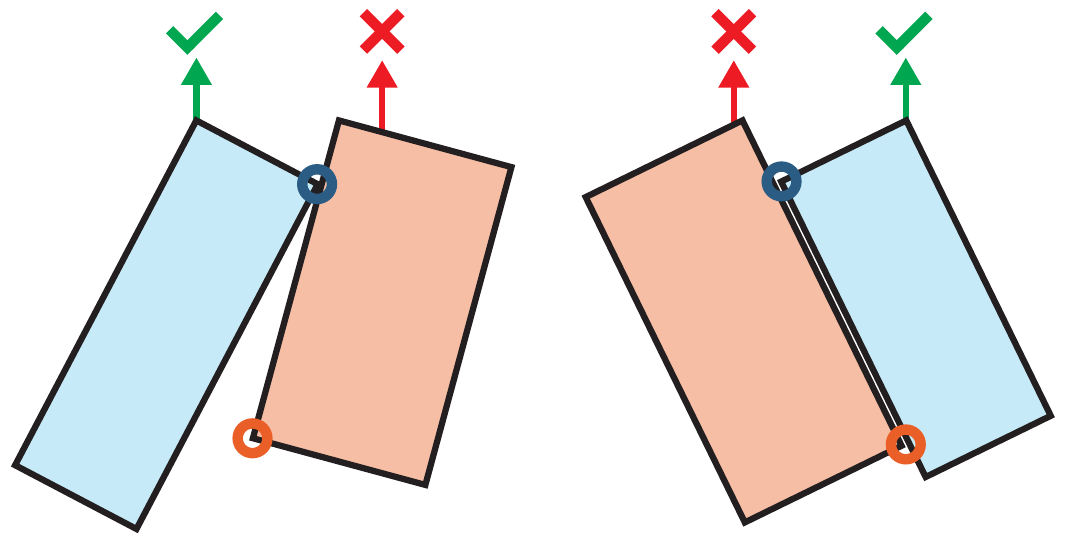}%
\end{wrapfigure}
If the OBB proxy was used, only one chart in each pair can potentially trigger intersection if moved up. We detect these cases by comparing the y-positions of the rightmost corner of the left OBB and the leftmost corner of the right OBB (circled in inset); whichever corner is lower corresponds to the chart whose upward movement may trigger intersection.
\end{parWithWrapFigure}

\paragraph*{Parallel Pushing Algorithm.} 
\label{sec:pushing_details}

\begin{figure}
\includegraphics[width=\linewidth]{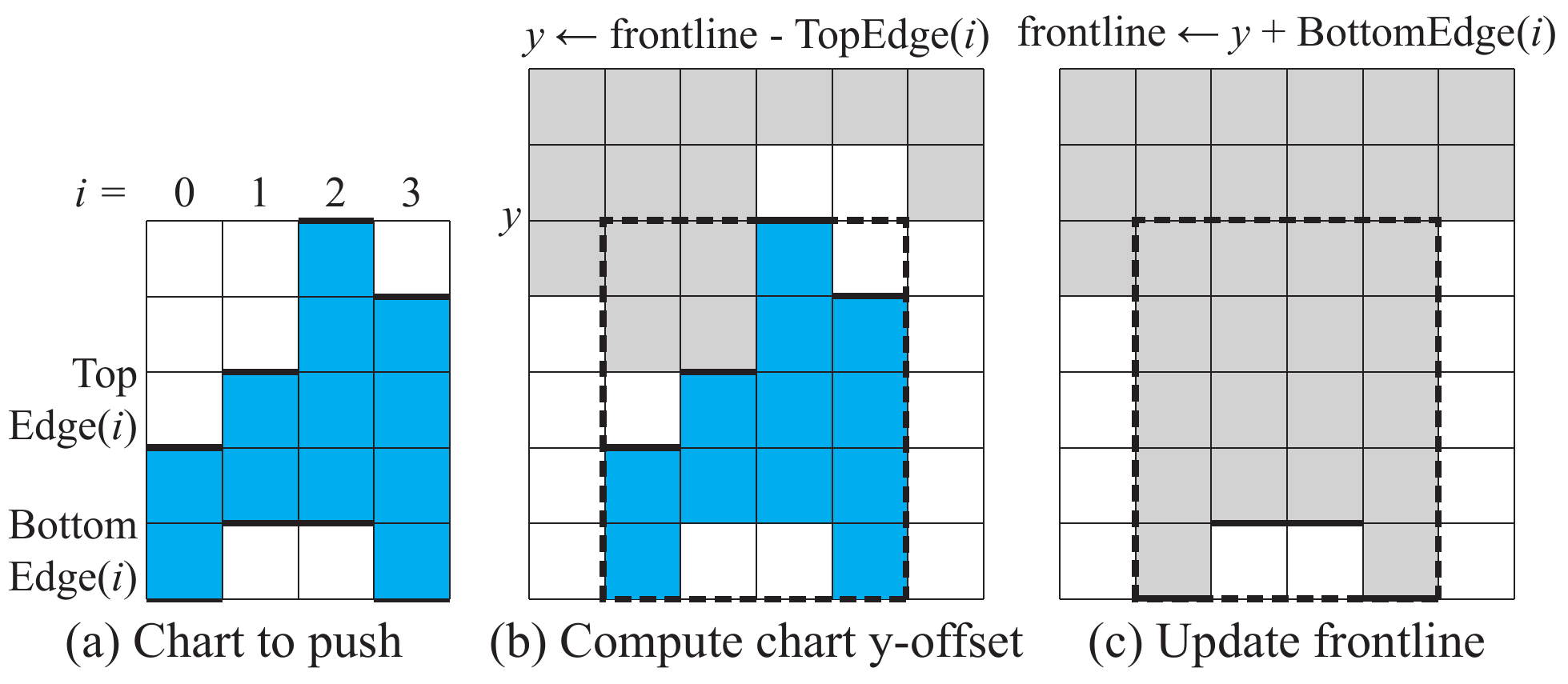}
\caption{Our extended GPU pushing algorithm. We define two functions for each chart, TopEdge and BottomEdge, which are computed by rounding the top and bottom boundaries of our chart shape proxies (a). In the first pass, each chart computes its vertical offset by reading all frontline texels it covers and comparing them to its TopEdge (b). Once chart vertical offsets are finalized, each chart updates the frontline with its BottomEdge (c).}
\label{fig:pushing}
\end{figure}

We provide full algorithmic details of our parallel pushing method below. All texels in the frontline are initialized with a height of zero, corresponding to the top of the atlas. We assume each chart defines two functions TopEdge($i$) and BottomEdge($i$), whose input $i$ is a integer texel index within $[0, \text{chartWidth})$ and whose output is an integer offset from the top of the chart in $[0, \text{chartHeight}]$. We process each texel in each chart independently; thus, the total number of processed texels is the sum of AABB widths of all charts in the row. First, each chart is pushed up to the existing frontline by setting its vertical offset to the maximum of (frontline $-$ TopEdge($i$)) across all texels that it covers horizontally. We then run Algorithm \ref{alg:interlock_correction} to eliminate all potential overlaps (refer to the previous section for the definition of \textit{CannotMoveAbove}). Once all charts' vertical offsets are finalized, each frontline texel that a chart covers horizontally is updated with the chart's vertical offset plus BottomEdge($i$); each frontline texel receives the maximum value across all charts that cover it. The process for one chart is illustrated in Fig. \ref{fig:pushing}.

We employ our two chart shape proxies to compute accurate values of TopEdge and BottomEdge at each texel.
For our local AABB proxy, we define TopEdge($i$) as the topmost $y$ position among the interval(s) in the piecewise constant top boundary that overlap texel $i$ horizontally; we define BottomEdge similarly. For our OBB proxy, we define TopEdge($i$) as the topmost $y$ position reached by the one or two ``top'' edges of the OBB within texel $i$; we define BottomEdge similarly. To strictly enforce chart separation and gutter requirements, TopEdge is rounded up to the nearest integer coordinate, while BottomEdge is rounded down. We evaluate TopEdge and BottomEdge using both proxies and use the result per-texel that provides the tightest bound on the chart shape; this corresponds to the larger $y$ coordinate for TopEdge and the smaller $y$ coordinate for BottomEdge.

Pseudocode for our full packing procedure is provided in Algorithms \ref{alg:folding} and \ref{alg:full_packing}.

\begin{algorithm}
\caption{Correcting chart vertical offsets after pushing}
\label{alg:interlock_correction}
\KwIn{array of per-chart offsets for the current scale factor}
\SetKwRepeat{Do}{do}{while}

\SetKwProg{Fn}{function}{}{end}
\Fn{CorrectYOffsets(offsets)}{
\Do{violationDetected}{
	violationDetected $\gets$ 0\;

	\ForPar{each pair of charts $(c, c + 1)$} {
		\tcc{offset is 0 at the top of the atlas, and increases going downward}
		\If{CannotMoveAbove($c$, $c + 1$) \textbf{and} offsets[$c$].y < offsets[$c + 1$].y}{
			atomicCompSwap(violationDetected, 0, 1)\;
			atomicMax(offsets[$c$].y, offsets[$c + 1$].y)\;
		}

		\If{CannotMoveAbove($c + 1$, $c$) \textbf{and} offsets[$c + 1$].y < offsets[$c$].y}{
			atomicCompSwap(violationDetected, 0, 1)\;
			atomicMax(offsets[$c + 1$].y, offsets[$c$].y)\;
		}
	}
}
}
\end{algorithm}

\paragraph*{Knee Detection Parameters.} \label{sec:knee_param_details}
We categorize the height difference between two consecutive charts as a knee if this difference is at least 10 percent of the atlas height and at least 20\% of the height of the taller chart, which we find to produce reasonable results in practice.
Because we consider knees a product of a row's shape and not necessarily its size, we use charts' original, unscaled heights in this comparison.

\begin{figure}
\includegraphics[width=\linewidth]{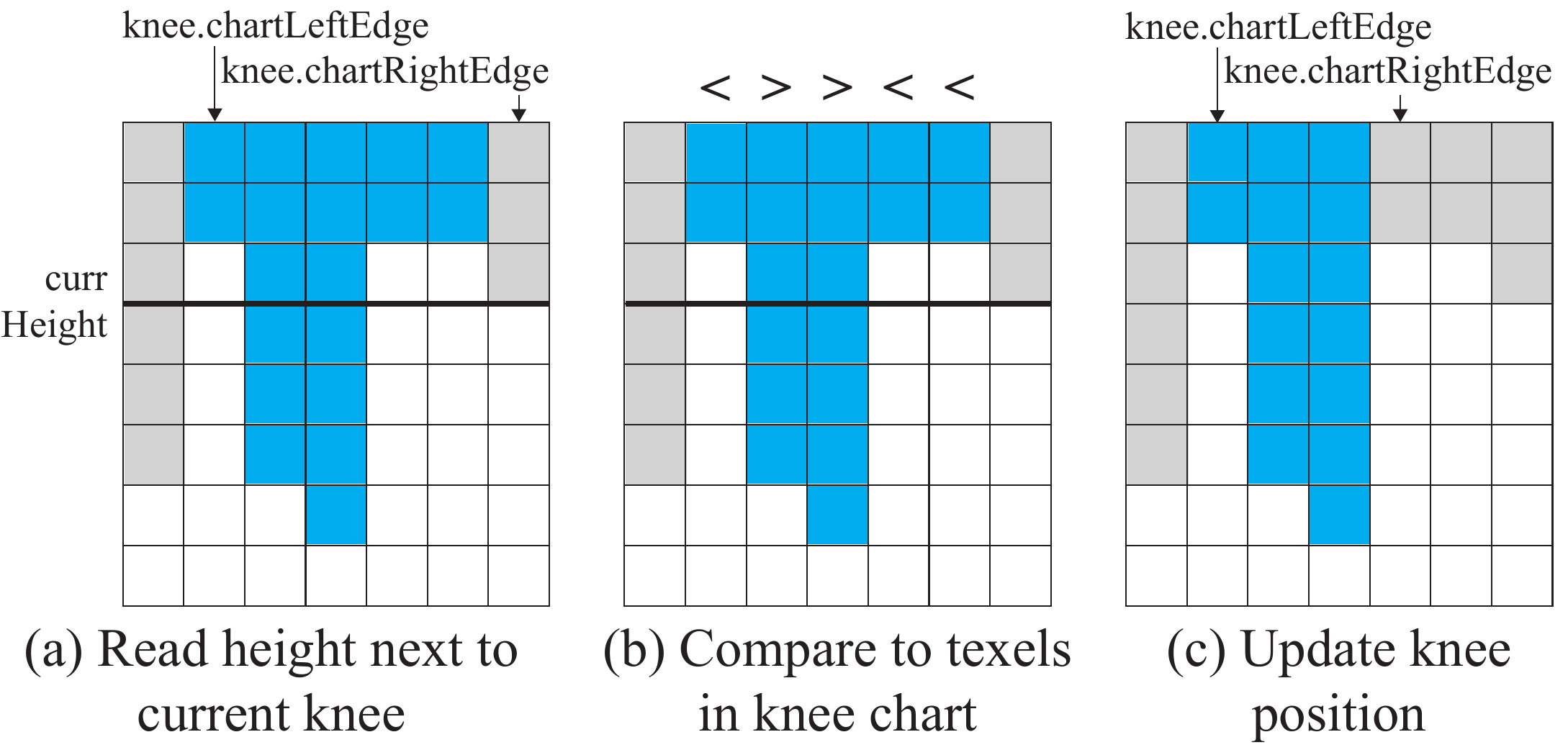}
\caption{Our parallel algorithm for computing the knee position from the frontline. Here a left-to-right row is illustrated; the right-to-left case is symmetric. We maintain the current left edge (inclusive) and right edge (exclusive) of the knee chart, and query the frontline height at the right edge (a). We find the closest intersection of the corresponding horizontal line with the knee chart by processing each frontline texel in the knee chart in parallel (b). Finally, we update the knee position and right edge of the knee chart using the location of the intersection (c).}
\label{fig:knee_position_update}
\end{figure}

\paragraph*{Updating Knee Location.} \label{sec:knee_loc_details}
Pseudocode for updating the location of the knee in the atlas is provided in Algorithm \ref{alg:knee_location}, and we illustrate the procedure in Fig \ref{fig:knee_position_update}.

\begin{algorithm}
\caption{Refining the knee location using the frontline}
\label{alg:knee_location}
\SetKwProg{Fn}{function}{}{end}
\Fn{UpdateKneeLocation(knee)}{
\eIf{knee.leftToRight}{
	currKneeLocation $\gets$ knee.chartRightEdge\;
	currHeight $\gets$ frontLine[currKneeLocation]\;
	newKneeLocation $\gets$ knee.chartLeftEdge - 1\;
	\ForPar{each texel $x$ in [knee.chartLeftEdge, knee.chartRightEdge)}{
		\If{frontLine[$x$] >= currHeight}{
			atomicMax(newKneeLocation, $x$)\;
		}
	}
	knee.chartRightEdge $\gets$ newKneeLocation + 1\;
}{
	currKneeLocation $\gets$ knee.chartLeftEdge - 1\;
	currHeight $\gets$ frontLine[currKneeLocation]\;
	newKneeLocation $\gets$ knee.chartRightEdge\;
	\ForPar{each texel $x$ in [knee.chartLeftEdge, knee.chartRightEdge)}{
		\If{frontLine[$x$] >= currHeight}{
			atomicMin(newKneeLocation, $x$)\;
		}
	}
	knee.chartLeftEdge $\gets$ newKneeLocation\;
}
}
\end{algorithm}

\begin{algorithm}
\caption{Folding charts into a row}
\label{alg:folding}
\KwIn{index of first chart in row; folding width; whether to enable horizontal compacting; per-chart offsets for the current scale factor}
\KwOut{index of last chart in row}

\SetKwProg{Fn}{function}{}{end}
\Fn{FoldRow(rowStart, foldingWidth, enableHC, offsets)}{

$c$ $\gets$ rowStart\;
currLeftEdge $\gets$ 0\;

\While{$c$ < numCharts}{
	nextLeftEdge $\gets$ currLeftEdge + GetChartWidth($c$)\;
	\If{nextLeftEdge > foldingWidth}{
		\Return{$c - 1$}\;
	}
	offsets[$c$].x $\gets$ currLeftEdge\;
	\eIf{enableHC}{
		currLeftEdge $\gets$ nextLeftEdge - compactingDistance[$c + 1$]\;
	}{
		currLeftEdge $\gets$ nextLeftEdge\;
	}
	$c$ $\gets$ $c + 1$\;
}
\Return{numCharts $- 1$}\;
}
\end{algorithm}

\begin{algorithm}
\caption{Full packing computation for one scale factor}
\label{alg:full_packing}
\KwOut{True if packing succeeded; false otherwise}

rowStart $\gets$ 0\;
overflow $\gets$ false\;
kneeData $\gets$ NULL\;
\tcc{while more rows can be formed}
\While{rowStart < numCharts $\land$ $\neg$overflow}{
	\If{kneeData $\neq$ NULL} {
		UpdateKneeLocation(kneeData)\;
	}
	\For{each configuration $j$ = (rowDirection, foldingWidth, enableHC)}{
		\tcc{folding step}
		rowEnd[$j$] $\gets$ FoldRow(rowStart, foldingWidth, enableHC, localOffsets[$j$])\;
		\If{rowDirection is right-to-left}{
			Reflect charts and chart x offsets\;
		}
		\tcc{pushing step}
		Copy frontLine to localFrontLine[$j$]\;
		\ForPar{each chart $c$, texel $i$ in chart}{
			$x$ $\gets$ localOffsets[$j$][$c$].x + $i$\;
			atomicMax(localOffsets[$j$][$c$].y, localFrontLine[$j$][$x$] - TopEdge$(c, i)$)\;
		}
		CorrectYOffsets(localOffsets[$j$])\;
		\ForPar{each chart $c$, texel $i$ in chart} {
			$x$ $\gets$ localOffsets[$j$][$c$].x + $i$\;
			atomicMax(localFrontLine[$j$][$x$], localOffsets[$j$][$c$].y + BottomEdge$(c, i)$)\;
		}
		\tcc{compute score}
		\ForPar{each texel $x$ in atlasWidth} {
			atomicMax(score[$j$], localFrontLine[$j$][$x$])\;
		}
		\If{configuration folded at knee}{
			\ForPar{each texel $x$ between atlas edge and knee location} {
				atomicMax(scoreKnee[$j$], localFrontLine[$j$][$x$])\;
			}
		}
	}

	\tcc{commit the best configuration}
	Set $j$ to best configuration\;

	\If{best configuration not folded at knee}{
		\tcc{check if this row contains a knee; if so, save its data}
		kneeData $\gets$ FindKnee(rowStart, rowEnd[$j$])\;
	}

	Copy localFrontLine[$j$] to frontLine\;
	Copy localOffsets[$j$][rowStart..rowEnd[$j$]] to offsets[rowStart..rowEnd[$j$]]\;
	overflow $\gets$ score[$j$] > atlasHeight\;
	rowStart $\gets$ rowEnd[$j$] + 1\;
}
\Return $\neg$overflow;
\end{algorithm}

\section{Additional Evaluation Details.}
\label{sec:data_details}

\paragraph*{Data Set Sources.}
We assemble our dataset of UV-unwrapped chart sets from the datasets of \cite{liu2017seamless}, \cite{knodt2025texture}, and \cite{yang2023learningbased}, and our dataset of TSS chart sets from that of \cite{vining2025fastatlas}. We exclude inputs with only 1 chart, which can be packed trivially, from both datasets. For UV chart sets with associated textures, we obtain charts' input coordinates in texel space, and thus their target size, by scaling the original UV coordinates in $[0,1]$ by the dimensions of the texture. Furthermore, we determine the ``full'' atlas dimensions (which should be square and a power of two) by taking the maximum side length in case of non-square textures and rounding down to a power of two.

We note that for TSS applications, where charts are often the result of clipping and projecting scene geometry into screen space, we are not aware of a clipping-free approximation of the OBB or of local AABBs akin to the conservative estimator used for AABBs in \cite{vining2025fastatlas}.
We therefore clip our TSS charts to the view frustum before passing them as input to any packing method.
However, software clipping is already required by some TSS algorithms (e.g., \cite{Neff2022MSA}); in this case, there will be no additional cost to clipping.

\paragraph*{UVPackMaster Evaluation Details.}

We configure UVPackMaster with the same packing settings used to evaluate all other methods: we set the \verb|pixelMargin| parameter to 2 pixels to reflect our target of 2 pixels' distance between charts, enable reflection and multiple-of-$90^\circ$ rotations, and enable pre-rotation to tight bounding boxes for UV charts only. Since UVPackMaster does not support searching for an optimal downscaling factor (which is at most $1$), we run a binary search over the 64 scale factors in the same way as for Xatlas.

On certain inputs containing complex chart topology, we find that UVPackMaster hangs indefinitely during the ``topology analysis'' step and does not terminate. All other methods, including ours, handle such inputs robustly. We detect these hangs by setting a timeout of 15 minutes; 639 of the 16,772 atlases we pack (4\%) exceed this timeout.

\begin{figure}
\includegraphics[width=\linewidth]{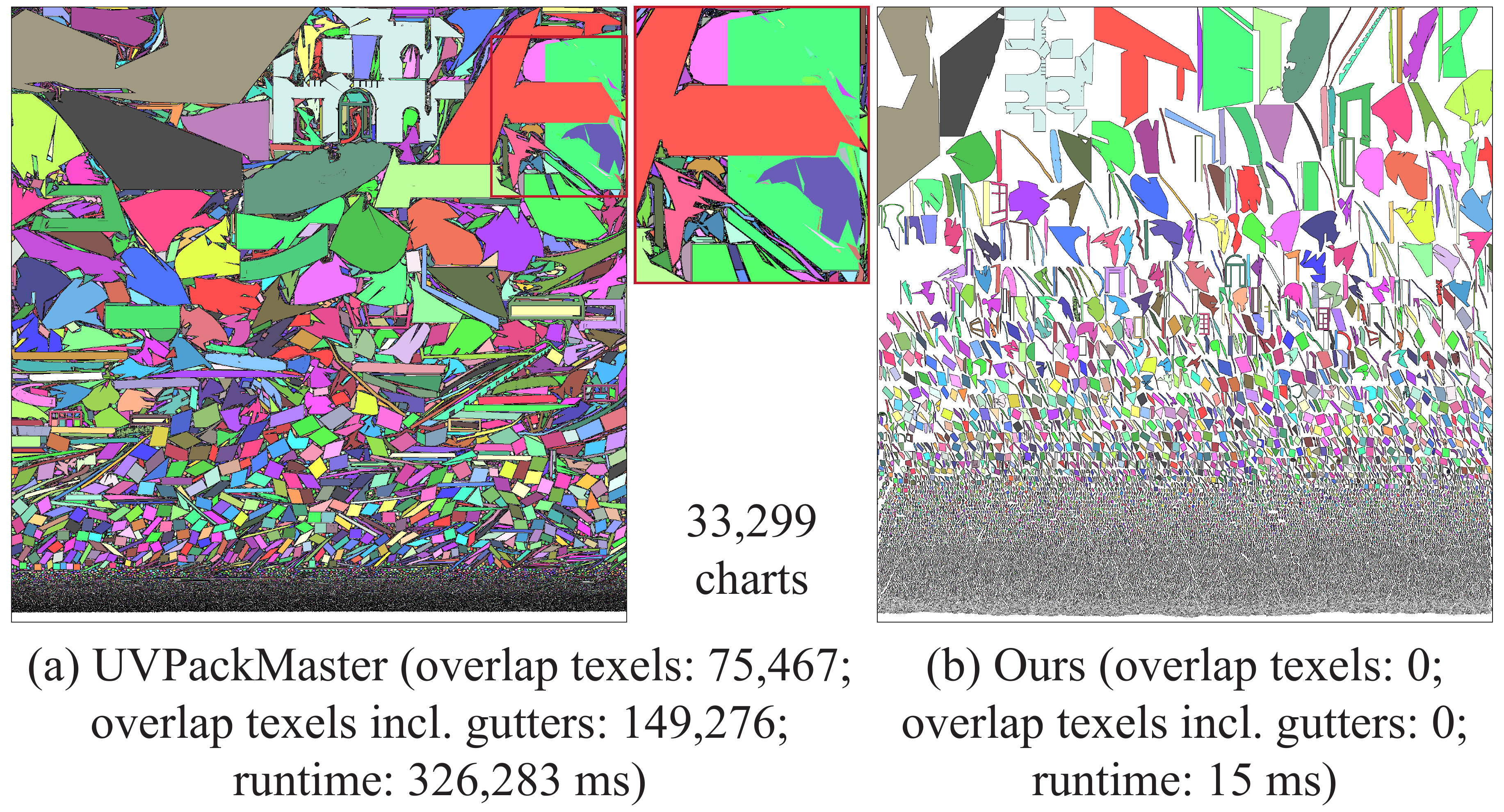}
\caption{
The commercial GPU-based packer UVPackMaster outputs packings with severe overlaps (a). Overlap size is measured as the number of texels covered by two or more charts at the same time. In addition, it fails to satisfy gutter requirements, measured by the number of texels covered by two or more charts when rendered with 1 pixel gutter dilation. Our method (b) is overlap-free and over 20,000 times faster.}
\label{fig:uvpm_overlap_SanMiguel}
\end{figure}

We set the \verb|precision| parameter to 2000 in order to eliminate chart overlap warnings raised by UVPackMaster at lower precision values. Nevertheless, even with these settings, UVPackMaster produces 1,202 atlases with severe overlaps for which no warnings are issued (7.5\% of inputs that terminate). Fig. \ref{fig:uvpm_overlap_SanMiguel} shows an additional example of a UVPackMaster output with such overlaps.

\section{Additional Gallery.}

\begin{figure*}
\includegraphics[width=\linewidth]{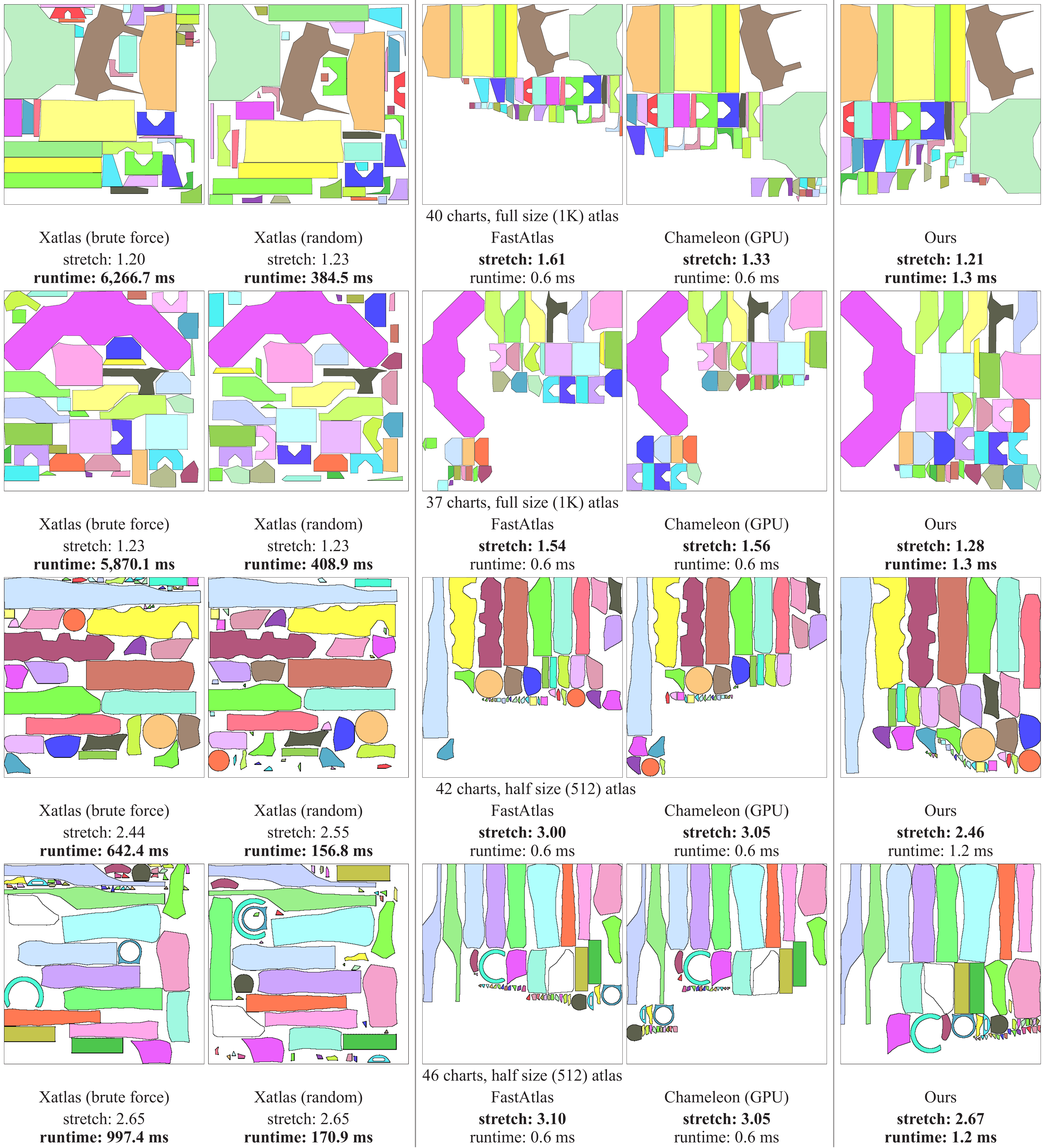}
\caption{Additional comparisons of our outputs against prior offline (Xatlas \cite{xatlas}) and interactive (FastAtlas \cite{vining2025fastatlas}, Chameleon \cite{igarashi2001adaptive}) alternatives on chart sets originating from UV unwrapping.}
\label{fig:results_uv_supp_1}
\end{figure*}

\begin{figure*}
\includegraphics[width=\linewidth]{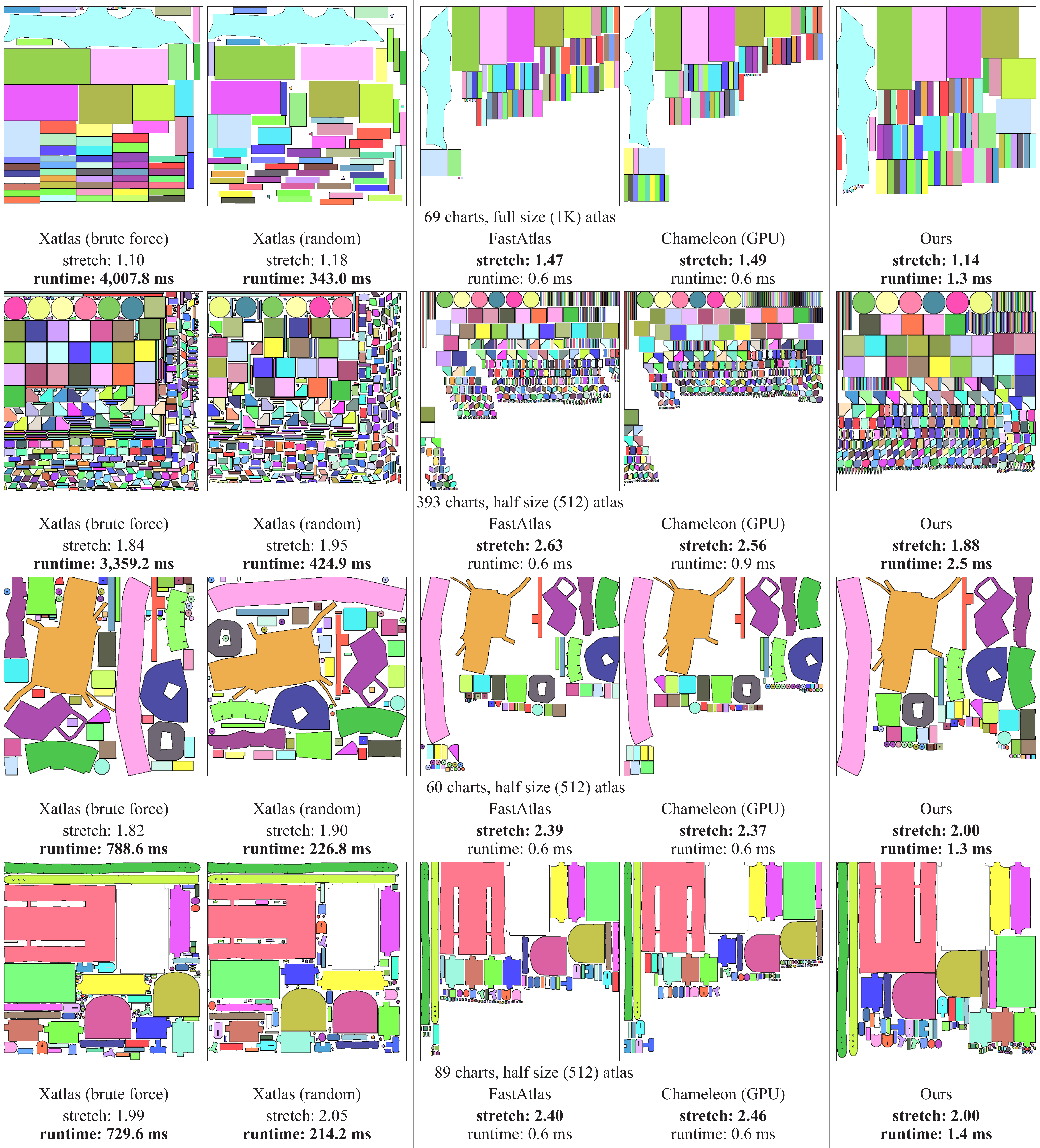}
\caption{Additional comparisons of our outputs against prior offline (Xatlas \cite{xatlas}) and interactive (FastAtlas \cite{vining2025fastatlas}, Chameleon \cite{igarashi2001adaptive}) alternatives on chart sets originating from UV unwrapping.}
\label{fig:results_uv_supp_2}
\end{figure*}

\begin{figure*}
\includegraphics[width=\linewidth]{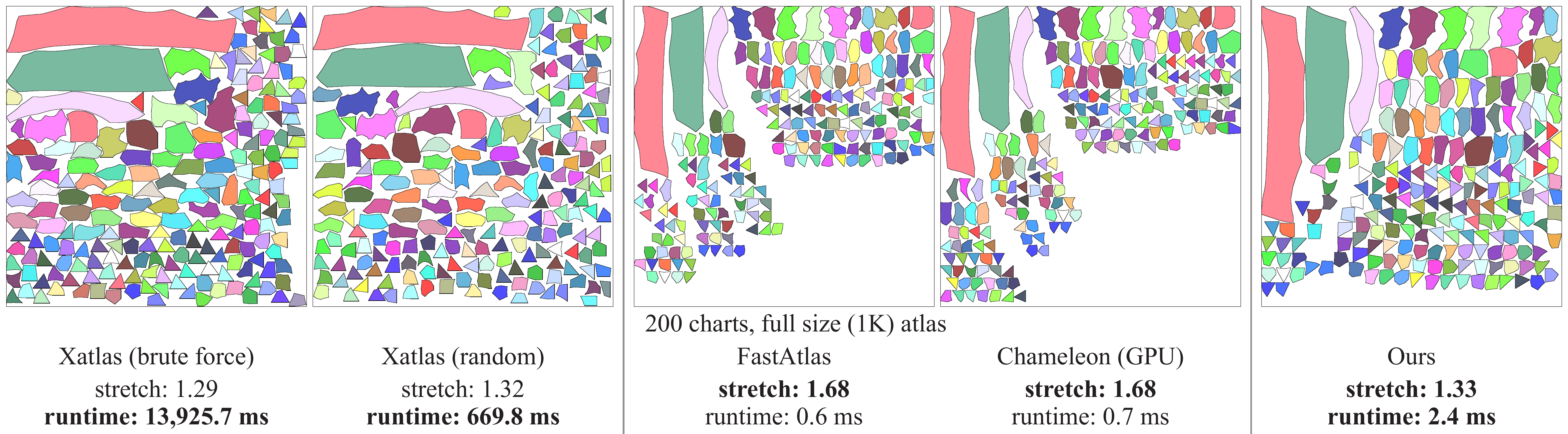}
\caption{Additional comparisons of our outputs against prior offline (Xatlas \cite{xatlas}) and interactive (FastAtlas \cite{vining2025fastatlas}, Chameleon \cite{igarashi2001adaptive}) alternatives on chart sets originating from UV unwrapping.}
\label{fig:results_uv_supp_3}
\end{figure*}

\begin{figure*}
\includegraphics[width=\linewidth]{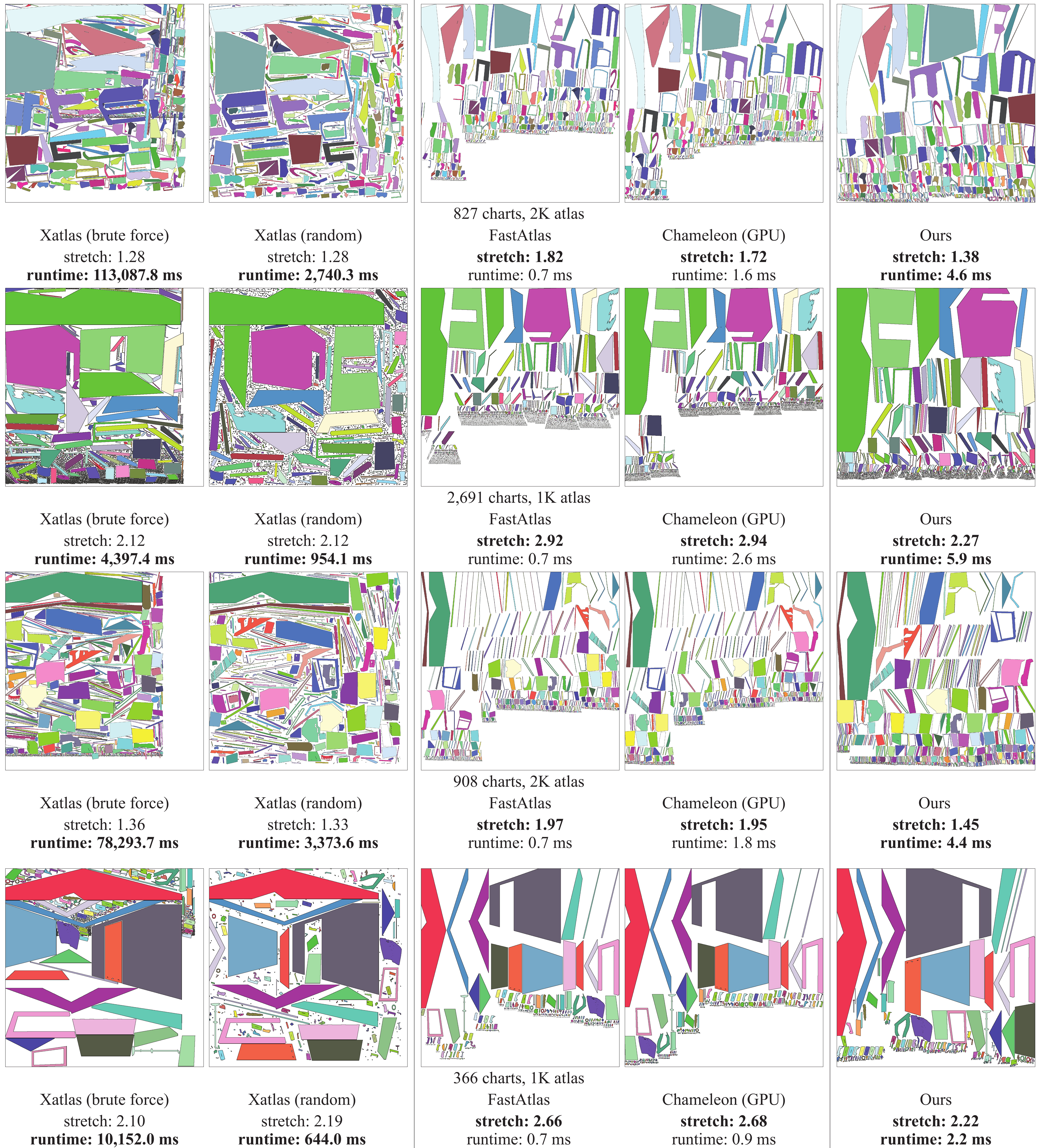}
\caption{Additional comparisons of our outputs against prior offline (Xatlas \cite{xatlas}) and interactive (FastAtlas \cite{vining2025fastatlas}, Chameleon \cite{igarashi2001adaptive}) alternatives on texture-space shading chart sets \cite{vining2025fastatlas}. For TSS data, charts are not pre-rotated to their tight bounding boxes.}
\label{fig:results_tss_supp}
\end{figure*}

Figs. \ref{fig:results_uv_supp_1}-\ref{fig:results_tss_supp} provide additional examples of atlases packed by our method compared to prior offline and interactive alternatives for both UV-unwrapped and texture-space shading charts.


\section{Additional Ablations and Statistics}

\paragraph*{Runtime Breakdown.}

Table \ref{tab:runtime_breakdown} reports the contribution of each step of our method to our overall runtime. As expected, folding and pushing across all candidate scale factors using our balanced strategy and vertical compacting (Alg. \ref{alg:full_packing}) occupies much of this time (approximately 60\%). Computing our chart shape proxies comprises approximately a further 20\%.

\begin{table}
\scriptsize
\setlength{\tabcolsep}{2pt}
\centering
\begin{tabular}{lc|cccccc|c}
                   &                    &                &                &               & \textbf{Local}     &                     &                  &                  \\
                   &                    &                &                &               & \textbf{AABB}      &                     &                  &                  \\
                   &                    & \textbf{AABB/} & \textbf{Local} &               & \textbf{Refine \&} & \textbf{Horizontal} &                  &                  \\
                   &                    & \textbf{OBB}   & \textbf{AABB}  & \textbf{AABB} & \textbf{Chart}     & \textbf{Distance}   & \textbf{Fold \&} & \textbf{Total}   \\
\textbf{\# charts} & \textbf{\# inputs} & \textbf{Comp.} & \textbf{Comp.} & \textbf{Sort} & \textbf{Orient.}   & \textbf{Comp.}      & \textbf{Push}    & \textbf{Packing} \\
\hline
$\leq$ 100         & 1,726              & 0.04           & 0.04           & 0.42          & 0.20               & 0.09                & 0.42             & 1.22             \\
101-500            & 2,338              & 0.06           & 0.10           & 0.43          & 0.40               & 0.18                & 0.95             & 2.13             \\
501-1,000          & 1,640              & 0.08           & 0.17           & 0.43          & 0.88               & 0.32                & 1.70             & 3.58             \\
1,001-5,000        & 8,508              & 0.10           & 0.29           & 0.50          & 1.16               & 0.44                & 2.83             & 5.31             \\
5,001-10,000       & 814                & 0.23           & 0.65           & 0.53          & 1.20               & 0.52                & 7.54             & 10.68            \\
$>$ 10,000         & 1,746              & 0.34           & 1.24           & 0.46          & 1.73               & 0.64                & 6.38             & 10.79            \\
\textbf{All}       & \textbf{16,772}    & \textbf{0.12}  & \textbf{0.34}  & \textbf{0.47} & \textbf{0.99}      & \textbf{0.38}       & \textbf{2.81}    & \textbf{5.11}   
\end{tabular}
\caption{Detailed runtime breakdown for each step of our method, averaged across all datasets and atlas sizes (all times in milliseconds). Left to right: AABB and OBB computation, local AABB computation, chart sorting by AABB height, boundary refinement for local AABBs and chart orientation computation, horizontal compacting distance computation, balanced folding and pushing for all candidate scale factors, and total packing time (sum of all previous columns).}
\label{tab:runtime_breakdown}
\vspace{-3mm}
\end{table}

\paragraph*{Additional Statistics.}

We provide detailed statistics for our experiments broken down by chart set source (UV charts, where charts are per-rotated to their tight boxes for all methods, and TSS charts, where they are not) and atlas size. Tables \ref{tab:runtime_uv}, \ref{tab:runtime_tss}, \ref{tab:stretch_uv}, and \ref{tab:stretch_tss} provide per-dataset and atlas size breakdowns of runtime and stretch for our method and alternatives. Tables \ref{tab:ablationBalancedTight_uv} and \ref{tab:ablationBalancedTight_tss} show the effect of isolating our tightness improvement or balance improvement components compared to our full method, broken down along the same lines. Since only 6 UV chart sets contain more than 500 charts, we do not create separate buckets for chart counts greater than 500 in our UV chart tables. Table \ref{tab:optimization_tss} shows the effect of our optimization strategy and $t_{opt}$ parameter on texture-space shading inputs with more than 1,000 charts, broken down by atlas size. Note that since only 2 UV chart sets exceed the 1,000-chart threshold, we are unable to include a similar breakdown for UV inputs.

\begin{table}
\scriptsize
\setlength{\tabcolsep}{2pt}
\centering
\begin{tabular}{lc|cc|cc|c|ccc}
& & \multicolumn{5}{c|}{Mean packing time (ms) $\downarrow$} & \multicolumn{3}{c}{Packing time ratio} \\
                   &                    &                   & \textbf{Xatlas}    &                & \textbf{Cham-} &               & \textbf{Xatlas}   & \textbf{Fast-}   & \textbf{Cham-}   \\
                   &                    & \textbf{Xatlas}   & \textbf{(brute}    & \textbf{Fast-} & \textbf{eleon} &               & \textbf{(random)} & \textbf{Atlas}   & \textbf{eleon}   \\
\textbf{\# charts} & \textbf{\# inputs} & \textbf{(random)} & \textbf{force)}    & \textbf{Atlas} & \textbf{(GPU)} & \textbf{Ours} & \textbf{to ours}  & \textbf{to ours} & \textbf{to ours} \\
\hline
\textbf{Half Size} &                    &                   &                    &                &                &               &                   &                  &                  \\
$\leq$ 100         & 364                & 183.68            & 2,118.82           & 0.61           & 0.59           & 1.12          & 156.19            & 0.55             & 0.53             \\
$>$ 100            & 91                 & 483.84            & 9,603.26           & 0.65           & 0.83           & 1.74          & 255.79            & 0.39             & 0.48             \\
\textbf{All}       & \textbf{455}       & \textbf{243.71}   & \textbf{3,615.71}  & \textbf{0.62}  & \textbf{0.64}  & \textbf{1.25} & \textbf{176.11}   & \textbf{0.52}    & \textbf{0.52}    \\
\hline
\textbf{Full Size} &                    &                   &                    &                &                &               &                   &                  &                  \\
$\leq$ 100         & 364                & 471.92            & 18,911.39          & 0.61           & 0.58           & 1.20          & 379.91            & 0.52             & 0.49             \\
$>$ 100            & 91                 & 966.28            & 46,617.56          & 0.64           & 0.79           & 1.90          & 468.66            & 0.35             & 0.42             \\
\textbf{All}       & \textbf{455}       & \textbf{570.79}   & \textbf{24,452.62} & \textbf{0.61}  & \textbf{0.62}  & \textbf{1.34} & \textbf{397.66}   & \textbf{0.48}    & \textbf{0.48}   
\end{tabular}
\caption{Runtime comparison of our method against both prior offline packing methods (Xatlas \cite{xatlas}) and alternatives capable of interactive performance (FastAtlas \cite{vining2025fastatlas}, our GPU implementation of Chameleon \cite{igarashi2001adaptive}) on UV charts at full and half atlas sizes. The last 3 columns show the average ratios between the runtime of alternatives and that of our method. Our method is orders of magnitude faster than prior offline methods, and though slower than FastAtlas and Chameleon on average, it remains interactive.}
\label{tab:runtime_uv}
\vspace{-3mm}
\end{table}

\begin{table}
\scriptsize
\setlength{\tabcolsep}{2pt}
\centering
\begin{tabular}{lc|cc|cc|c|ccc}
& & \multicolumn{5}{c|}{Mean packing time (ms) $\downarrow$} & \multicolumn{3}{c}{Packing time ratio} \\
                   &                    &                   & \textbf{Xatlas}     &                & \textbf{Cham-} &               & \textbf{Xatlas}   & \textbf{Fast-}   & \textbf{Cham-}   \\
                   &                    & \textbf{Xatlas}   & \textbf{(brute}     & \textbf{Fast-} & \textbf{eleon} &               & \textbf{(random)} & \textbf{Atlas}   & \textbf{eleon}   \\
\textbf{\# charts} & \textbf{\# frames} & \textbf{(random)} & \textbf{force)}     & \textbf{Atlas} & \textbf{(GPU)} & \textbf{Ours} & \textbf{to ours}  & \textbf{to ours} & \textbf{to ours} \\
\hline
\textbf{1K}        &                    &                   &                     &                &                &               &                   &                  &                  \\
$\leq$ 100         & 499                & 441.63            & 6,502.50            & 0.63           & 0.61           & 1.18          & 372.11            & 0.57             & 0.56             \\
101-500            & 1,084              & 723.66            & 10,928.50           & 0.66           & 0.83           & 1.99          & 372.48            & 0.35             & 0.42             \\
501-1,000          & 816                & 969.13            & 10,219.32           & 0.69           & 1.28           & 3.33          & 292.27            & 0.21             & 0.39             \\
1,001-5,000        & 4,252              & 1,262.79          & 12,124.44           & 0.83           & 2.43           & 4.99          & 257.28            & 0.17             & 0.48             \\
5,001-10,000       & 407                & 2,671.16          & 23,592.42           & 1.42           & 7.36           & 10.04         & 265.25            & 0.14             & 0.73             \\
$>$ 10,000         & 873                & 5,675.60          & 51,512.06           & 1.64           & 16.34          & 9.85          & 577.87            & 0.17             & 1.63             \\
\textbf{All}       & \textbf{7,931}     & \textbf{1,665.23} & \textbf{16,335.32}  & \textbf{0.90}  & \textbf{3.76}  & \textbf{4.96} & \textbf{319.55}   & \textbf{0.22}    & \textbf{0.61}    \\
\hline
\textbf{2K}        &                    &                   &                     &                &                &               &                   &                  &                  \\
$\leq$ 100         & 499                & 1,271.39          & 65,283.53           & 0.64           & 0.61           & 1.36          & 941.02            & 0.52             & 0.50             \\
101-500            & 1,084              & 2,021.61          & 136,571.84          & 0.68           & 0.86           & 2.33          & 893.23            & 0.30             & 0.38             \\
501-1,000          & 816                & 2,538.68          & 122,868.79          & 0.72           & 1.51           & 3.85          & 660.95            & 0.19             & 0.39             \\
1,001-5,000        & 4,252              & 3,000.36          & 144,730.90          & 0.87           & 3.09           & 5.63          & 547.93            & 0.16             & 0.54             \\
5,001-10,000       & 407                & 4,280.75          & 136,668.11          & 1.48           & 9.31           & 11.32         & 381.30            & 0.13             & 0.82             \\
$>$ 10,000         & 873                & 6,921.68          & 93,189.72           & 1.73           & 19.91          & 11.73         & 601.53            & 0.15             & 1.69             \\
\textbf{All}       & \textbf{7,931}     & \textbf{3,207.64} & \textbf{130,280.62} & \textbf{0.94}  & \textbf{4.64}  & \textbf{5.69} & \textbf{628.83}   & \textbf{0.20}    & \textbf{0.64}   
\end{tabular}
\caption{Runtime comparison of our method against both prior offline packing methods (Xatlas \cite{xatlas}) and alternatives capable of interactive performance (FastAtlas \cite{vining2025fastatlas}, our GPU implementation of Chameleon \cite{igarashi2001adaptive}) on texture-space shading charts at 1K and 2K atlas sizes. The last 3 columns show the average ratios between the runtime of alternatives and that of our method. Our method is orders of magnitude faster than prior offline methods, and though slower than FastAtlas and Chameleon on average, it remains interactive.}
\label{tab:runtime_tss}
\vspace{-3mm}
\end{table}

\begin{table}
\scriptsize
\setlength{\tabcolsep}{2pt}
\centering
\begin{tabular}{lc|cc|cc|c|cc}
& & \multicolumn{5}{c|}{Mean stretch $\downarrow$} & \multicolumn{2}{c}{Stretch improvement} \\
                   &                    &                   & \textbf{Xatlas} &                    & \textbf{Cham-} &               &                    & \textbf{vs.}   \\
                   &                    & \textbf{Xatlas}   & \textbf{(brute} &                    & \textbf{eleon} &               & \textbf{vs.}       & \textbf{Cham-} \\
\textbf{\# charts} & \textbf{\# inputs} & \textbf{(random)} & \textbf{force)} & \textbf{FastAtlas} & \textbf{(GPU)} & \textbf{Ours} & \textbf{FastAtlas} & \textbf{eleon} \\
\hline
\textbf{Half Size} &                    &                   &                 &                    &                &               &                    & \textbf{}      \\
$\leq$ 100         & 364                & 2.58              & 2.51            & 2.81               & 2.71           & 2.60          & 60\%               & 43\%           \\
$>$ 100            & 91                 & 2.44              & 2.38            & 2.78               & 2.68           & 2.51          & 65\%               & 50\%           \\
\textbf{All}       & \textbf{455}       & \textbf{2.55}     & \textbf{2.49}   & \textbf{2.80}      & \textbf{2.70}  & \textbf{2.58} & \textbf{61\%}      & \textbf{44\%}  \\
\hline
\textbf{Full Size} &                    &                   &                 &                    &                &               &                    &                \\
$\leq$ 100         & 364                & 1.26              & 1.23            & 1.38               & 1.33           & 1.27          & 62\%               & 51\%           \\
$>$ 100            & 91                 & 1.19              & 1.16            & 1.35               & 1.31           & 1.24          & 64\%               & 46\%           \\
\textbf{All}       & \textbf{455}       & \textbf{1.25}     & \textbf{1.22}   & \textbf{1.37}      & \textbf{1.32}  & \textbf{1.26} & \textbf{62\%}      & \textbf{50\%} 
\end{tabular}
\caption{Packing scale comparison of our method against both prior offline packing methods (Xatlas \cite{xatlas}) and alternatives capable of interactive performance (FastAtlas \cite{vining2025fastatlas}, our GPU implementation of Chameleon \cite{igarashi2001adaptive}) on UV charts at full and half atlas sizes. Our method consistently outperforms interactive alternatives in terms of packing scale (average $L^2$ stretch between packed and input charts). The last two columns show the amount by which our method closes the stretch gap between interactive alternatives and the best method per-input.}
\label{tab:stretch_uv}
\vspace{-3mm}
\end{table}

\begin{table}
\scriptsize
\setlength{\tabcolsep}{2pt}
\centering
\begin{tabular}{lc|cc|cc|c|cc}
& & \multicolumn{5}{c|}{Mean stretch $\downarrow$} & \multicolumn{2}{c}{Stretch improvement} \\
                   &                    &                   & \textbf{Xatlas} &                    & \textbf{Cham-} &               &                    & \textbf{vs.}   \\
                   &                    & \textbf{Xatlas}   & \textbf{(brute} &                    & \textbf{eleon} &               & \textbf{vs.}       & \textbf{Cham-} \\
\textbf{\# charts} & \textbf{\# frames} & \textbf{(random)} & \textbf{force)} & \textbf{FastAtlas} & \textbf{(GPU)} & \textbf{Ours} & \textbf{FastAtlas} & \textbf{eleon} \\
\hline
\textbf{1K}        &                    &                   &                 &                    &                &               &                    & \textbf{}      \\
$\leq$ 100         & 499                & 1.89              & 1.85            & 2.16               & 2.13           & 2.01          & 36\%               & 33\%           \\
101-500            & 1,084              & 2.25              & 2.21            & 2.85               & 2.72           & 2.51          & 51\%               & 40\%           \\
501-1,000          & 816                & 2.36              & 2.32            & 3.28               & 3.21           & 2.71          & 56\%               & 52\%           \\
1,001-5,000        & 4,252              & 2.36              & 2.37            & 3.42               & 3.35           & 2.87          & 51\%               & 48\%           \\
5,001-10,000       & 407                & 2.76              & 3.49            & 4.28               & 4.04           & 3.40          & 58\%               & 50\%           \\
$>$ 10,000         & 873                & 4.76              & 6.15            & 5.79               & 5.26           & 4.47          & 76\%               & 68\%           \\
\textbf{All}       & \textbf{7,931}     & \textbf{2.60}     & \textbf{2.78}   & \textbf{3.56}      & \textbf{3.42}  & \textbf{2.95} & \textbf{54\%}      & \textbf{49\%}  \\
\hline
\textbf{2K}        &                    &                   &                 &                    &                &               &                    &                \\
$\leq$ 100         & 499                & 1.02              & 1.02            & 1.12               & 1.09           & 1.05          & 57\%               & 48\%           \\
101-500            & 1,084              & 1.12              & 1.11            & 1.40               & 1.34           & 1.24          & 54\%               & 41\%           \\
501-1,000          & 816                & 1.16              & 1.13            & 1.58               & 1.55           & 1.32          & 55\%               & 53\%           \\
1,001-5,000        & 4,252              & 1.15              & 1.14            & 1.62               & 1.61           & 1.38          & 49\%               & 48\%           \\
5,001-10,000       & 407                & 1.23              & 1.34            & 1.86               & 1.82           & 1.56          & 47\%               & 44\%           \\
$>$ 10,000         & 873                & 1.45              & 2.06            & 2.18               & 2.11           & 1.81          & 49\%               & 43\%           \\
\textbf{All}       & \textbf{7,931}     & \textbf{1.18}     & \textbf{1.24}   & \textbf{1.63}      & \textbf{1.60}  & \textbf{1.39} & \textbf{51\%}      & \textbf{47\%} 
\end{tabular}
\caption{Packing scale comparison of our method against both prior offline packing methods (Xatlas \cite{xatlas}) and alternatives capable of interactive performance (FastAtlas \cite{vining2025fastatlas}, our GPU implementation of Chameleon \cite{igarashi2001adaptive}) on texture-space shading charts at 1K and 2K atlas sizes. Our method consistently outperforms interactive alternatives in terms of packing scale (average $L^2$ stretch between packed and input charts). The last two columns show the amount by which our method closes the stretch gap between interactive alternatives and the best method per-frame.}
\label{tab:stretch_tss}
\vspace{-3mm}
\end{table}

\begin{table}
\scriptsize
\setlength{\tabcolsep}{2pt}
\centering
\begin{tabular}{lc|ccc|ccc}
& & \multicolumn{3}{c|}{Mean packing time (ms) $\downarrow$} & \multicolumn{3}{c}{Mean stretch $\downarrow$} \\
                   &                    & \textbf{Balanced} & \textbf{Tight} &               & \textbf{Balanced} & \textbf{Tight} &               \\
\textbf{\# charts} & \textbf{\# inputs} & \textbf{only}     & \textbf{only}  & \textbf{Ours} & \textbf{only}     & \textbf{only}  & \textbf{Ours} \\
\hline
\textbf{Half Size} &                    &                   &                &               &                   &                &               \\
$\leq$ 100         & 364                & 0.62              & 0.91           & 1.12          & 2.65              & 2.65           & 2.60          \\
$>$ 100            & 91                 & 0.81              & 1.29           & 1.74          & 2.57              & 2.59           & 2.51          \\
\textbf{All}       & \textbf{455}       & \textbf{0.65}     & \textbf{0.98}  & \textbf{1.25} & \textbf{2.64}     & \textbf{2.64}  & \textbf{2.58} \\
\hline
\textbf{Full Size} &                    &                   &                &               &                   &                &               \\
$\leq$ 100         & 364                & 0.63              & 0.93           & 1.20          & 1.30              & 1.30           & 1.27          \\
$>$ 100            & 91                 & 0.85              & 1.34           & 1.90          & 1.26              & 1.27           & 1.24          \\
\textbf{All}       & \textbf{455}       & \textbf{0.67}     & \textbf{1.01}  & \textbf{1.34} & \textbf{1.29}     & \textbf{1.29}  & \textbf{1.26}
\end{tabular}
\caption{Quantitative comparison of our method against our balance component only and our tightness component only on UV charts at full and half atlas sizes (left: average packing time in milliseconds, right: average $L^2$ stretch between packed and input charts). Our full method, which combines both components, achieves the best packing scale.}
\label{tab:ablationBalancedTight_uv}
\vspace{-3mm}
\end{table}

\begin{table}
\scriptsize
\setlength{\tabcolsep}{2pt}
\centering
\begin{tabular}{lc|ccc|ccc}
& & \multicolumn{3}{c|}{Mean packing time (ms) $\downarrow$} & \multicolumn{3}{c}{Mean stretch $\downarrow$} \\
                   &                    & \textbf{Balanced} & \textbf{Tight} &               & \textbf{Balanced} & \textbf{Tight} &               \\
\textbf{\# charts} & \textbf{\# frames} & \textbf{only}     & \textbf{only}  & \textbf{Ours} & \textbf{only}     & \textbf{only}  & \textbf{Ours} \\
\hline
\textbf{1K}        &                    &                   &                &               &                   &                &               \\
$\leq$ 100         & 499                & 0.68              & 0.97           & 1.18          & 2.07              & 2.03           & 2.01          \\
101-500            & 1,084              & 0.89              & 1.54           & 1.99          & 2.64              & 2.57           & 2.51          \\
501-1,000          & 816                & 1.20              & 2.56           & 3.33          & 2.95              & 2.93           & 2.71          \\
1,001-5,000        & 4,252              & 1.96              & 3.92           & 4.99          & 3.11              & 3.03           & 2.87          \\
5,001-10,000       & 407                & 4.86              & 7.81           & 10.04         & 3.71              & 3.61           & 3.40          \\
$>$ 10,000         & 873                & 3.27              & 8.19           & 9.85          & 4.95              & 4.75           & 4.47          \\
\textbf{All}       & \textbf{7,931}     & \textbf{1.95}     & \textbf{3.94}  & \textbf{4.96} & \textbf{3.20}     & \textbf{3.11}  & \textbf{2.95} \\
\hline
\textbf{2K}        &                    &                   &                &               &                   &                &               \\
$\leq$ 100         & 499                & 0.71              & 0.99           & 1.36          & 1.07              & 1.05           & 1.05          \\
101-500            & 1,084              & 0.97              & 1.60           & 2.33          & 1.31              & 1.27           & 1.24          \\
501-1,000          & 816                & 1.33              & 2.63           & 3.85          & 1.44              & 1.42           & 1.32          \\
1,001-5,000        & 4,252              & 2.14              & 4.01           & 5.63          & 1.50              & 1.46           & 1.38          \\
5,001-10,000       & 407                & 5.17              & 8.04           & 11.32         & 1.70              & 1.65           & 1.56          \\
$>$ 10,000         & 873                & 3.70              & 9.02           & 11.73         & 1.97              & 1.92           & 1.81          \\
\textbf{All}       & \textbf{7,931}     & \textbf{2.13}     & \textbf{4.11}  & \textbf{5.69} & \textbf{1.50}     & \textbf{1.46}  & \textbf{1.39}
\end{tabular}
\caption{Quantitative comparison of our method against our balance component only and our tightness component only on texture-space shading charts at 1K and 2K atlas sizes (left: average packing time in milliseconds, right: average $L^2$ stretch between packed and input charts). Our full method, which combines both components, achieves the best packing scale.}
\label{tab:ablationBalancedTight_tss}
\vspace{-3mm}
\end{table}

\begin{table}
\scriptsize
\setlength{\tabcolsep}{2pt}
\centering
\begin{tabular}{lc|cccc|c|cccc|c}
& & \multicolumn{5}{c|}{Mean packing time (ms) $\downarrow$} & \multicolumn{5}{c}{Mean stretch $\downarrow$} \\
                   &                    & \textbf{$\mathbf{t_{opt}}$} & \textbf{$\mathbf{t_{opt}}$} & \textbf{$\mathbf{t_{opt}}$} & \textbf{$\mathbf{t_{opt}}$} &               & \textbf{$\mathbf{t_{opt}}$} & \textbf{$\mathbf{t_{opt}}$} & \textbf{$\mathbf{t_{opt}}$} & \textbf{$\mathbf{t_{opt}}$} &                      \\
\textbf{\# charts} & \textbf{\# frames} & \textbf{= 5\%}              & \textbf{= 2\%}              & \textbf{= 1\%}              & \textbf{= 0\%}              & \textbf{Ours} & \textbf{= 5\%}              & \textbf{= 2\%}              & \textbf{= 1\%}              & \textbf{= 0\%}              & \textbf{Ours}        \\
\hline
\textbf{1K}        &                    &                             &                             &                             &                             &               &                             &                             &                             & \multicolumn{1}{l}{}        & \multicolumn{1}{l}{} \\
1,001-5,000        & 4,252              & 4.26                        & 4.27                        & 4.34                        & 4.99                        & 4.99          & 2.89                        & 2.88                        & 2.88                        & 2.87                        & 2.87                 \\
5,001-10,000       & 407                & 5.69                        & 5.82                        & 6.28                        & 10.04                       & 10.04         & 3.50                        & 3.47                        & 3.46                        & 3.40                        & 3.40                 \\
$>$ 10,000         & 873                & 8.45                        & 8.82                        & 9.85                        & 19.48                       & 9.85          & 4.56                        & 4.52                        & 4.47                        & 4.24                        & 4.47                 \\
\textbf{All}       & \textbf{5,532}     & \textbf{5.02}               & \textbf{5.10}               & \textbf{5.35}               & \textbf{7.65}               & \textbf{6.13} & \textbf{3.20}               & \textbf{3.18}               & \textbf{3.17}               & \textbf{3.12}               & \textbf{3.16}        \\
\hline
\textbf{2K}        &                    &                             &                             &                             &                             &               &                             &                             &                             &                             &                      \\
1,001-5,000        & 4,252              & 4.99                        & 4.98                        & 5.06                        & 5.63                        & 5.63          & 1.39                        & 1.38                        & 1.38                        & 1.38                        & 1.38                 \\
5,001-10,000       & 407                & 6.87                        & 6.99                        & 7.56                        & 11.32                       & 11.32         & 1.58                        & 1.57                        & 1.57                        & 1.56                        & 1.56                 \\
$>$ 10,000         & 873                & 10.18                       & 10.62                       & 11.73                       & 21.72                       & 11.73         & 1.84                        & 1.82                        & 1.81                        & 1.78                        & 1.81                 \\
\textbf{All}       & \textbf{5,532}     & \textbf{5.94}               & \textbf{6.02}               & \textbf{6.30}               & \textbf{8.59}               & \textbf{7.01} & \textbf{1.47}               & \textbf{1.47}               & \textbf{1.46}               & \textbf{1.46}               & \textbf{1.46}       
\end{tabular}
\caption{Quantitative evaluation of different values of $t_{opt}$ on texture-space shading frames with more than 1,000 charts at 1K and 2K atlas sizes (left: average packing time in milliseconds, right: average $L^2$ stretch between packed and input charts). Higher values of $t_{opt}$ reduce packing time, but increase stretch. Our method, which sets $t_{opt}=0\%$ for inputs with up to 10,000 charts and $t_{opt}=1\%$ otherwise, achieves similar packing scale to $t_{opt}=0\%$ while retaining interactive average runtimes.}
\label{tab:optimization_tss}
\vspace{-3mm}
\end{table}

\paragraph*{Local AABB Count Ablation.}

We ablate the impact of the number of local AABBs by comparing the runtime and stretch of our method with 2 and 5 local AABBs per axis to our chosen value of 10 (Tab. \ref{tab:ablationSubBoxes}). As expected, the fewer local AABBs we allocate, the lower the runtime but the more the stretch increases. Tables \ref{tab:ablationSubBoxes_uv} and \ref{tab:ablationSubBoxes_tss} provide detailed breakdowns by dataset and atlas size.

\begin{table}
\scriptsize
\setlength{\tabcolsep}{2pt}
\centering
\begin{tabular}{lc|ccc|ccc}
& & \multicolumn{3}{c|}{Mean packing time (ms) $\downarrow$} & \multicolumn{3}{c}{Mean stretch $\downarrow$} \\
                   &                    & \textbf{Ours}      & \textbf{Ours}      &               & \textbf{Ours}      & \textbf{Ours}      &               \\
\textbf{\# charts} & \textbf{\# inputs} & \textbf{(2 boxes)} & \textbf{(5 boxes)} & \textbf{Ours} & \textbf{(2 boxes)} & \textbf{(5 boxes)} & \textbf{Ours} \\
\hline
$\leq$ 100         & 1,726              & 0.80               & 0.93               & 1.22          & 1.71               & 1.70               & 1.70          \\
101-500            & 2,338              & 1.16               & 1.45               & 2.13          & 1.93               & 1.90               & 1.88          \\
501-1,000          & 1,640              & 1.70               & 2.29               & 3.58          & 2.10               & 2.04               & 2.01          \\
1,001-5,000        & 8,508              & 2.71               & 3.49               & 5.31          & 2.20               & 2.15               & 2.12          \\
5,001-10,000       & 814                & 6.34               & 7.71               & 10.68         & 2.56               & 2.50               & 2.48          \\
$>$ 10,000         & 1,746              & 5.08               & 6.64               & 10.79         & 3.22               & 3.18               & 3.14          \\
\textbf{All}       & \textbf{16,772}    & \textbf{2.62}      & \textbf{3.36}      & \textbf{5.11} & \textbf{2.23}      & \textbf{2.18}      & \textbf{2.16}
\end{tabular}
\caption{Quantitative comparisons of our method with different numbers of local AABBs: 2, 5, and our full method (10) averaged across all datasets and atlas sizes (left: average packing time in milliseconds, right: average $L^2$ stretch between packed and input charts). Packing scale improves and the packing time increases with the number of local AABBs, providing a means to control the performance-quality tradeoff.}
\label{tab:ablationSubBoxes}
\vspace{-3mm}
\end{table}

\begin{table}
\scriptsize
\setlength{\tabcolsep}{2pt}
\centering
\begin{tabular}{lc|ccc|ccc}
& & \multicolumn{3}{c|}{Mean packing time (ms) $\downarrow$} & \multicolumn{3}{c}{Mean stretch $\downarrow$} \\
                   &                    & \textbf{Ours}      & \textbf{Ours}      &               & \textbf{Ours}      & \textbf{Ours}      &               \\
\textbf{\# charts} & \textbf{\# inputs} & \textbf{(2 boxes)} & \textbf{(5 boxes)} & \textbf{Ours} & \textbf{(2 boxes)} & \textbf{(5 boxes)} & \textbf{Ours} \\
\hline
\textbf{Half Size} &                    &                    &                    &               &                    &                    &               \\
$\leq$ 100         & 364                & 0.74               & 0.86               & 1.12          & 2.62               & 2.60               & 2.60          \\
$>$ 100            & 91                 & 1.00               & 1.23               & 1.74          & 2.54               & 2.53               & 2.51          \\
\textbf{All}       & \textbf{455}       & \textbf{0.79}      & \textbf{0.93}      & \textbf{1.25} & \textbf{2.60}      & \textbf{2.59}      & \textbf{2.58} \\
\hline
\textbf{Full Size} &                    &                    &                    &               &                    &                    &               \\
$\leq$ 100         & 364                & 0.77               & 0.91               & 1.20          & 1.28               & 1.28               & 1.27          \\
$>$ 100            & 91                 & 1.07               & 1.33               & 1.90          & 1.25               & 1.24               & 1.24          \\
\textbf{All}       & \textbf{455}       & \textbf{0.83}      & \textbf{1.00}      & \textbf{1.34} & \textbf{1.28}      & \textbf{1.27}      & \textbf{1.26}
\end{tabular}
\caption{Quantitative comparisons of our method with different numbers of local AABBs: 2, 5, and our full method (10) on UV charts at full and half atlas sizes (left: average packing time in milliseconds, right: average $L^2$ stretch between packed and input charts). Packing scale improves and the packing time increases with the number of local AABBs, providing a means to control the performance-quality tradeoff.}
\label{tab:ablationSubBoxes_uv}
\vspace{-3mm}
\end{table}

\begin{table}
\scriptsize
\setlength{\tabcolsep}{2pt}
\centering
\begin{tabular}{lc|ccc|ccc}
& & \multicolumn{3}{c|}{Mean packing time (ms) $\downarrow$} & \multicolumn{3}{c}{Mean stretch $\downarrow$} \\
                   &                    & \textbf{Ours}      & \textbf{Ours}      &               & \textbf{Ours}      & \textbf{Ours}      &               \\
\textbf{\# charts} & \textbf{\# frames} & \textbf{(2 boxes)} & \textbf{(5 boxes)} & \textbf{Ours} & \textbf{(2 boxes)} & \textbf{(5 boxes)} & \textbf{Ours} \\
\hline
\textbf{1K}        &                    &                    &                    &               &                    &                    &               \\
$\leq$ 100         & 499                & 0.80               & 0.92               & 1.18          & 2.03               & 2.01               & 2.01          \\
101-500            & 1,084              & 1.10               & 1.36               & 1.99          & 2.58               & 2.53               & 2.51          \\
501-1,000          & 816                & 1.59               & 2.11               & 3.33          & 2.83               & 2.75               & 2.71          \\
1,001-5,000        & 4,252              & 2.56               & 3.28               & 4.99          & 2.97               & 2.90               & 2.87          \\
5,001-10,000       & 407                & 6.07               & 7.31               & 10.04         & 3.50               & 3.43               & 3.40          \\
$>$ 10,000         & 873                & 4.65               & 6.00               & 9.85          & 4.57               & 4.51               & 4.47          \\
\textbf{All}       & \textbf{7,931}     & \textbf{2.56}      & \textbf{3.26}      & \textbf{4.96} & \textbf{3.05}      & \textbf{2.99}      & \textbf{2.95} \\
\hline
\textbf{2K}        &                    &                    &                    &               &                    &                    &               \\
$\leq$ 100         & 499                & 0.86               & 1.01               & 1.36          & 1.05               & 1.05               & 1.05          \\
101-500            & 1,084              & 1.25               & 1.57               & 2.33          & 1.28               & 1.25               & 1.24          \\
501-1,000          & 816                & 1.82               & 2.47               & 3.85          & 1.38               & 1.34               & 1.32          \\
1,001-5,000        & 4,252              & 2.85               & 3.70               & 5.63          & 1.43               & 1.40               & 1.38          \\
5,001-10,000       & 407                & 6.60               & 8.10               & 11.32         & 1.61               & 1.57               & 1.56          \\
$>$ 10,000         & 873                & 5.51               & 7.28               & 11.73         & 1.88               & 1.84               & 1.81          \\
\textbf{All}       & \textbf{7,931}     & \textbf{2.89}      & \textbf{3.74}      & \textbf{5.69} & \textbf{1.44}      & \textbf{1.41}      & \textbf{1.39}
\end{tabular}
\caption{Quantitative comparisons of our method with different numbers of local AABBs: 2, 5, and our full method (10) on texture-space shading charts at 1K and 2K atlas sizes (left: average packing time in milliseconds, right: average $L^2$ stretch between packed and input charts). Packing scale improves and the packing time increases with the number of local AABBs, providing a means to control the performance-quality tradeoff.}
\label{tab:ablationSubBoxes_tss}
\vspace{-3mm}
\end{table}

\end{document}